\documentclass[twocolumn]{aastex62}
\usepackage{graphicx}	
\usepackage[caption=false]{subfig}

\usepackage{amssymb}
\usepackage{color}
\usepackage{amsmath}
\usepackage{longtable}

\usepackage{natbib}
\newcommand{\zfourge}{{\sc zfourge}}
\newcommand{\Hbeta}{H$\beta$}
\newcommand{\OIII}{\hbox{{\rm [O}\kern 0.1em{\sc iii}{\rm ]}}}
\newcommand{\Ks}{$K_s$}
\newcommand{\Av}{A$_{\rm V}$}

\newcommand{\UVJ}{\textit{UVJ}}
\newcommand{\UVc}{\textit{(U-V)}}
\newcommand{\VJc}{\textit{(V-J)}}
\newcommand{\Dfour}{\textit{D}(4000)}

\begin{document}

\title{ZFOURGE: Using Composite Spectral Energy Distributions to Characterize Galaxy Populations at $1<\MakeLowercase{z}<4$\footnote{This Paper includes data gathered with the 6.5 m Magellan Telescopes located at Las Campanas Observatory, Chile.}}

\correspondingauthor{Ben Forrest}
\email{bforrest@physics.tamu.edu}

\author[0000-0001-6003-0541]{Ben Forrest}
\affiliation{George P. and Cynthia W. Mitchell Institute for Fundamental Physics and Astronomy, Department of Physics and Astronomy, Texas A\&M University, College Station, TX 77843, USA}

\author[0000-0001-9208-2143]{Kim-Vy H. Tran}
\affiliation{George P. and Cynthia W. Mitchell Institute for Fundamental Physics and Astronomy, Department of Physics and Astronomy, Texas A\&M University, College Station, TX 77843, USA}
\affiliation{School of Physics, University of New South Wales, Kensington, Australia}
\affiliation{Australian Astronomical Observatory, PO Box 915, North Ryde, NSW 1670, Australia}

\author{Adam Broussard}
\affiliation{Department of Physics and Astronomy, Rutgers, The State University of New Jersey, 136 Frelinghuysen Road, Piscataway, NJ 08854, USA}

\author[0000-0003-1420-6037]{Jonathan H. Cohn}
\affiliation{George P. and Cynthia W. Mitchell Institute for Fundamental Physics and Astronomy, Department of Physics and Astronomy, Texas A\&M University, College Station, TX 77843, USA}

\author{Robert C. Kennicutt, Jr.}
\affiliation{Steward Observatory, University of Arizona, 933 N Cherry Avenue, Tucson, AZ  85721-0065, USA}

\author[0000-0001-7503-8482]{Casey Papovich}
\affiliation{George P. and Cynthia W. Mitchell Institute for Fundamental Physics and Astronomy, Department of Physics and Astronomy, Texas A\&M University, College Station, TX 77843, USA}

\author[0000-0002-7278-9528]{Rebecca Allen}
\affiliation{Australian Astronomical Observatory, PO Box 915, North Ryde, NSW 1670, Australia}
\affiliation{Centre for Astrophysics and Supercomputing, Swinburne University, Hawthorn, VIC 3122, Australia}

\author[0000-0002-4653-8637]{Michael Cowley}
\affiliation{Australian Astronomical Observatory, PO Box 915, North Ryde, NSW 1670, Australia}
\affiliation{Department of Physics \& Astronomy, Macquarie University, Sydney, NSW 2109, Australia}

\author[0000-0002-3254-9044]{Karl Glazebrook}
\affiliation{Centre for Astrophysics and Supercomputing, Swinburne University, Hawthorn, VIC 3122, Australia}

\author{Glenn G. Kacprzak}
\affiliation{Centre for Astrophysics and Supercomputing, Swinburne University, Hawthorn, VIC 3122, Australia}

\author[0000-0003-4032-2445]{Lalitwadee Kawinwanichakij}
\affiliation{George P. and Cynthia W. Mitchell Institute for Fundamental Physics and Astronomy, Department of Physics and Astronomy, Texas A\&M University, College Station, TX 77843, USA}
\affiliation{LSSTC Data Science Fellow}

\author[0000-0003-2804-0648]{Themiya Nanayakkara}
\affiliation{Leiden Observatory, Niels Bohrweg 2, 2333 CA, Leiden, Netherlands}

\author[0000-0002-7453-7279]{Brett Salmon}
\affiliation{Space Telescope Science Institute, Baltimore, MD, USA}

\author[0000-0001-5185-9876]{Lee R. Spitler}
\affiliation{Australian Astronomical Observatory, PO Box 915, North Ryde, NSW 1670, Australia}
\affiliation{Department of Physics \& Astronomy, Macquarie University, Sydney, NSW 2109, Australia}

\author[0000-0001-5937-4590]{Caroline M. S. Straatman}
\affiliation{Sterrenkundig Observatorium, Universiteit Gent, Krijgslaan 281 S9, B-9000 Gent, Belgium}

\begin{abstract}
We investigate the properties of galaxies as they shut off star formation over the 4 billion years surrounding peak cosmic star formation.
To do this we categorize $\sim7000$ galaxies from $1<z<4$ into $90$ groups based on the shape of their spectral energy distributions (SEDs) and build composite  SEDs with $R\sim 50$ resolution.
These composite SEDs show a variety of spectral shapes and also show trends in parameters such as color, mass, star formation rate, and emission line equivalent width. 
Using emission line equivalent widths and strength of the 4000\AA\ break, \Dfour, we categorize the composite SEDs into five classes: extreme emission line, star-forming, transitioning, post-starburst, and quiescent galaxies.
The transitioning population of galaxies show modest H$\alpha$ emission ($EW_{\rm REST}\sim40$\AA) compared to more typical star-forming composite SEDs at $\log_{10}(M/M_\odot)\sim10.5$ ($EW_{\rm REST}\sim80$\AA).
Together with their smaller sizes (3 kpc vs. 4 kpc) and higher S\'ersic indices (2.7 vs. 1.5), this indicates that morphological changes initiate before the cessation of star formation.
The transitional group shows a strong increase of over one dex in number density from $z\sim3$ to $z\sim1$, similar to the growth in the quiescent population, while post-starburst galaxies become rarer at $z\lesssim1.5$.
We calculate average quenching timescales of 1.6 Gyr at $z\sim1.5$ and 0.9 Gyr at $z\sim2.5$ and conclude that a fast quenching mechanism producing post-starbursts dominated the quenching of galaxies at early times, while a slower process has become more common since $z\sim2$.

\end{abstract}

\section{Introduction}

Since the beginning of the millennium, the number of galaxies with multi-wavelength photometric observations and accurate redshifts has exploded.
A wide range of surveys including the Deep Lens Survey \citep{Wittman2002}, Sloan Digital Sky Survey \citep{Abazajian2003}, imaging in the Hawaii Hubble Deep Field North \citep{Capak2004}, the Newfirm Medium Band Survey \citep{vanDokkum2009}, \textit{3D-HST} \citep{vanDokkum2011}, the Cosmic Assembly Near-Infrared Deep Extragalactic Legacy Survey \citep[\textit{CANDELS};][]{Koekemoer2011}, and the FourStar Galaxy Evolution Survey \citep[\zfourge;][]{Straatman2016} have increased our knowledge of galaxy formation and evolution tremendously.
With upcoming facilities such as the Large Synoptic Survey Telescope (LSST), we will soon truly be in an era where analyzing each individual galaxy will be prohibitive.
As such, we must find automated ways to study large numbers of galaxies.
One approach is to group galaxies together based on common spectral characteristics-- optimizing this methodology will be an important piece of understanding the lifecycles of galaxies through cosmic time.

Previous studies have grouped galaxies together in a variety of ways.
Often galaxies with similar values of a given parameter e.g., mass, star formation rate (SFR), S\'ersic index, radius, rest-frame color, emission line strength, or infrared (IR) luminosity, will be analyzed together, and all such categorizations can tease out important pieces of information \citep[e.g.,][]{Shapley2003, Brinchmann2008a, Nakajima2014, Eales2017, Eales2017b}.
Perhaps most prevalent in extragalactic studies, plotting the rest-frame colors \UVc\ and \VJc\ against one another has been used to classify galaxies into star-forming or quiescent regimes, approximate dust content, and constrain galaxy evolution \citep[e.g.,][]{Labbe2005, Wuyts2007, Williams2009, Whitaker2011, Brammer2011, Patel2012, Papovich2015}.
More recently, other trends in this \UVJ\ diagram have been noticed for high redshift populations, such as increasing specific SFR perpendicular to the quiescent wedge \citep[see Figure 26 of][]{Straatman2016, Fang2018}.

More statistically robust methods have also been used for grouping galaxies, as early as in \cite{Miller1996} with the use of Self Organizing Maps.
More recent methods have included local linear embedding \citep{Vanderplas2009}, principal components analysis \citep[PCA;][]{Wild2014, Maltby2016, Rowlands2018}, and composite SED construction \citep{Kriek2011, Kriek2013a, Forrest2016, Forrest2017a}.
For the latter, using medium-band and broadband filters to construct composite SEDs allows for impressive sensitivity and sample size.
At the same time, this method enables analysis of emission lines and discriminates more clearly between stellar populations than is typically possible without spectroscopic data.

In this work, we spectral diagnostics calculated from composite SEDs to categorize galaxies and show that this classification scheme accurately picks out rare populations, as supported by other properties and scaling relations.
This includes galaxies with strong nebular emission lines (Emission Line Galaxies-- ELG), as well as galaxies transitioning from star-forming (SFGs) to quiescent (QGs) regimes, which we split into two groups-- transitional galaxies (TGs), which show H$\alpha$ emission, and post-starburst galaxies (PSBs), which do not show H$\alpha$ emission.

PSBs have been a historically rare population, and have been studied in small numbers for some time \citep[e.g.,][]{Couch1987, Tran2003, Tran2004, Poggianti2009}.
Such galaxies have recently undergone a period of strong star formation, which has stopped within the last several hundred million years.
As a result, their spectra are dominated by main sequence A stars with significant Balmer absorption \citep[e.g.,][]{Dressler1983}.
While analysis of these galaxies allows insight into the mechanisms by which galaxies cease forming stars, such galaxies generally require spectroscopic confirmation, further preventing large samples from being found, particularly at higher redshifts.
Additionally, it is not clear that all galaxies undergo such a phase, as the mechanisms behind the quenching of galaxies are still uncertain, and may vary \citep{Tran2003, Wilkinson2017}.

The timescale for which galaxies remain in this post-starburst state is thought to be on the order of $10^8$ years \citep[e.g.][]{Wild2016}, and may be dependent upon environment \citep[e.g.,][]{Tran2003, Tran2004, Poggianti2009}.
As this timescale is relatively short, finding such galaxies is somewhat challenging, and several methods have been used to more easily identify these objects.
\citet{Whitaker2012} use \UVJ\ selection and single stellar population models, while other recent works such as \citet{Wild2014, Wild2016} have used principal components analysis for identifying post-starburst galaxies from multi-wavelength photometry alone.
Spectroscopic follow-up of these objects \citep{Maltby2016} have shown a high success rate for this method.

Alternative pathways to quenching are also suggested by the population of non-PSB galaxies in what has come to be called the `green valley' introduced in \citet{Martin2007, Salim2007, Schiminovich2007, Wyder2007}-- in this work we use the term transitional galaxies.
Originally selected to be between the star-forming sequence and quenched population of the color-magnitude diagram, similar galaxies have since been selected based on relations between colors, stellar masses, stellar mass surface densities, and SFRs \citep[e.g.,][]{Mendez2011, Fang2013, Schawinski2014, Pandya2017}.
Studies have hypothesized different quenching routes that galaxies may take before shutting off star formation permanently, including the idea of rejuvenation, in which a galaxy stops and restarts star formation multiple times \citep[e.g.,][]{Darvish2016, Pandya2017, Nelson2017, Dave2017}.
Here we use composite SEDs to infer quenching timescales of galaxies over the 4 billion years around peak cosmic star formation.
 
This paper builds on the composite SED work published in \citet{Forrest2016, Forrest2017a}.
Here we reconstruct composite SEDs using the full \zfourge\ sample (previous work used a subset of the full dataset) and provide a more detailed description of our data and methodology in Sections \ref{Data} \& \ref{Methods}, respectively.
Section \ref{Measure} relays our measurements based on the composite SEDs, as well as parameters from the individual galaxies themselves.
We then present our composite SEDs in terms of spectral features from the composite SEDs and analysis of the photometry of individual galaxies (Sections \ref{SAnalysis} \& \ref{PAnalysis}).
Discussion of the TGs (Section \ref{Disc}) and conclusions (Section \ref{Conc}) follow.
The entire set of composite SEDs and associated parameters are presented in the Appendix.
Throughout the work we assume a cosmology with $H_0=70$ km s$^{-1}$ Mpc$^{-1}$, $\Omega_m=0.3$, and $\Omega_\Lambda=0.7$ and make use of the AB magnitude system.

\section{Data} \label{Data}

We use multi-wavelength photometry from the FourStar Galaxy Evolution Survey \citep[\zfourge;][]{Straatman2016} in our work.
This survey obtained deep near-IR imaging with the \texttt{FourStar} imager \citep{Persson2013} of three legacy fields: CDFS \citep{Giacconi2002}, COSMOS \citep{Scoville2007}, and UDS \citep{Lawrence2007}.
\citet{Straatman2016} combined $K$-band imaging data from a number of surveys \citep{Retzlaff2010, Hsieh2012, McCracken2012, Fontana2014, Almaini2017} to create deep mosaics used as the detection images for the \zfourge\ catalogs (see Section 2.3 of \citet{Straatman2016} for details).
Morphological data for \zfourge\ galaxies cross-matched with \textit{HST/WFC3/F160W} \textit{CANDELS} data from \citet{VanderWel2012} are also included.

In addition to these data, multi-wavelength data from a variety of sources were included in a set of publicly released catalogs \citep{Giavalisco2004, Erben2005, Hildebrandt2006, Taniguchi2007, Furusawa2008, Wuyts2008, Erben2008, Hildebrandt2009, Nonino2009, Cardamone2010, Grogin2011, Koekemoer2011, Windhorst2011, Brammer2012}.
The CDFS, COSMOS, and UDS fields have 40, 37, and 26 filter bandpass observations ranging from 0.3-8 $\mu$m with 80\% completeness limits of 26.0, 25.5, and 25.8 AB magnitudes in the stacked \Ks\ band, respectively \citep{Straatman2016}. 
These catalogs are particularly well suited to the composite SED method due to their accurate photometric redshifts \citep[$1-2$\%;][]{ Nanayakkara2016}, broad range of rest-frame wavelengths probed, and deep imaging which allows for inclusion of faint galaxies at high redshifts.

Star formation rates are from publicly available catalogs compiled by \citet{Tomczak2016}, which used legacy UV data as well as data from Spitzer/MIPS (GOODS-S: PI Dickinson, COSMOS: PI Scoville, UDS: PI Dunlop) and Herschel/PACS (GOODS-S: \citet{Elbaz2011}, COSMOS \& UDS: PI Dickinson).
AGN host catalogs from \citet{Cowley2016} are also provided in the \zfourge\ data release.

\section{Composite SED Construction} \label{Methods}

\subsection{Sample Selection}


	\begin{figure}[tp]
	\centerline{\includegraphics[width=0.5\textwidth,trim=0in 0in 0in 0in, clip=true]{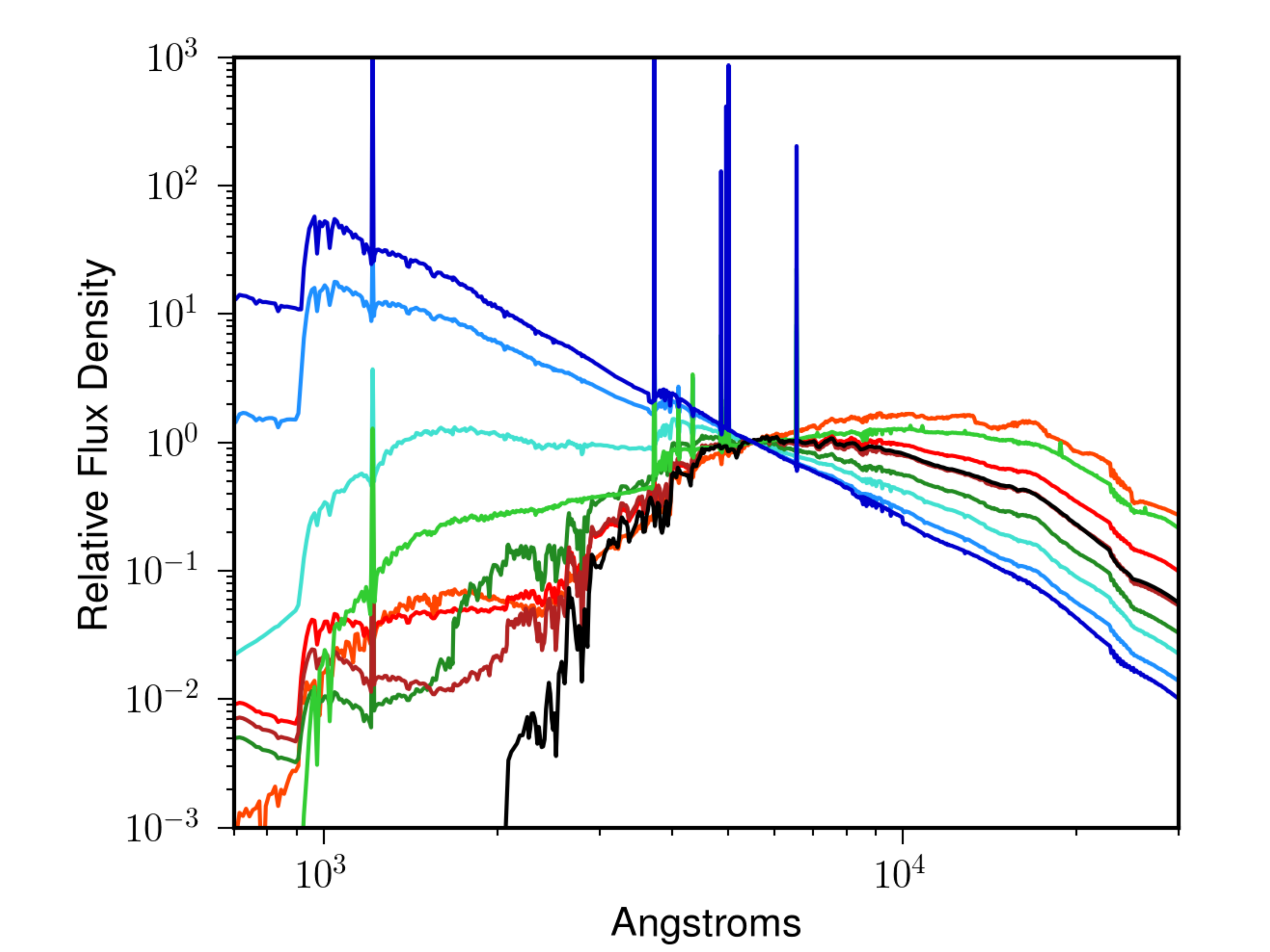}}
	\caption{EAZY templates used to fit galaxy SEDs, the same as used in \citet{Straatman2016}. The template with the greatest flux at 1000\AA\ is a high-EW model from \citet{Erb2010}, while the template with the greatest flux at 1$\mu$m is an old, dusty template. Other templates are included with EAZY \citep{Brammer2008}.}
	\label{fig:EAZY}
	\end{figure}



	\begin{figure*}[tp]
	\centerline{\includegraphics[width=\textwidth,trim=0in 0in 0in 0in, clip=true]{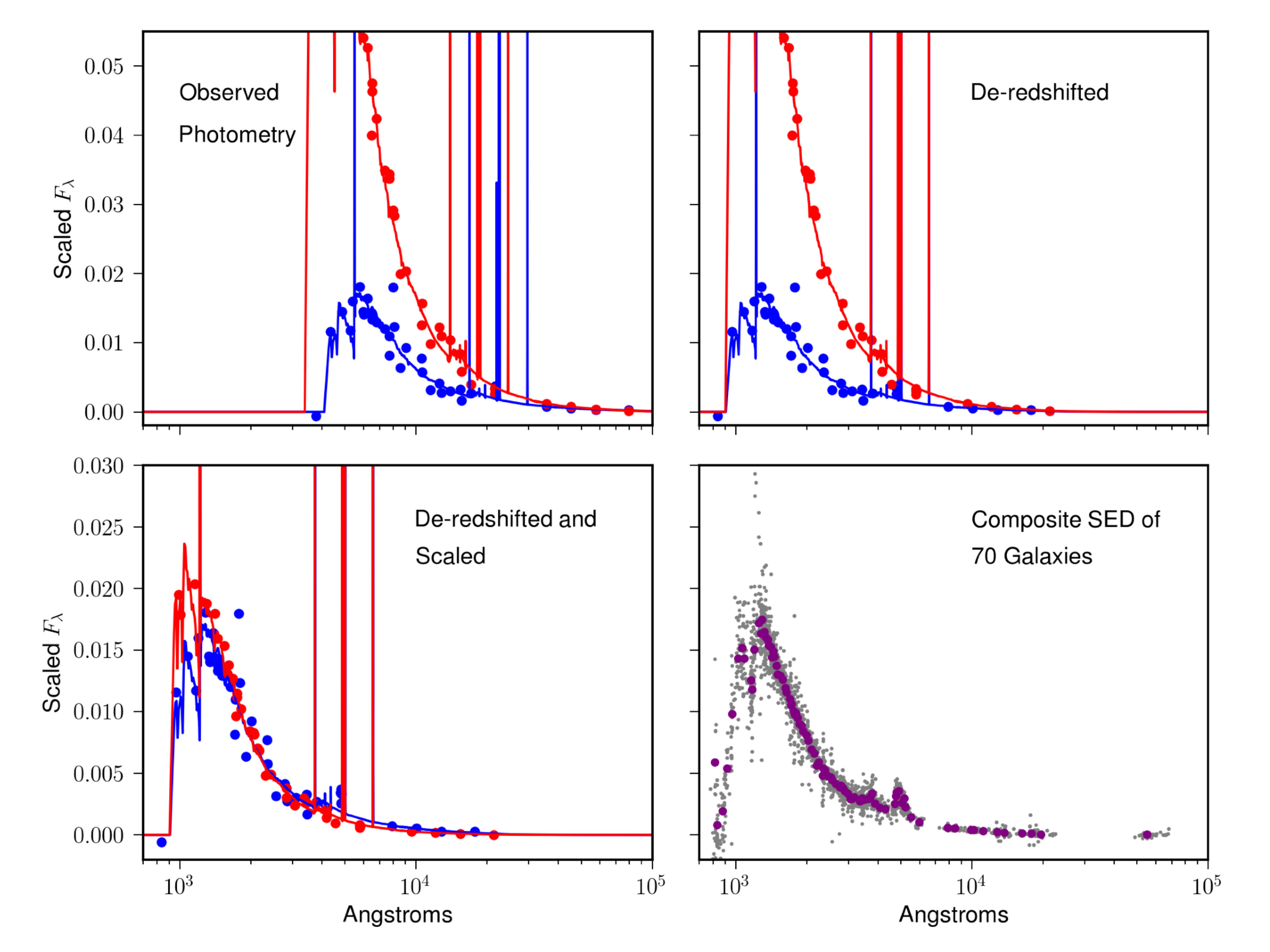}}
	\caption{Basic method of composite SED construction. The observed photometry (points) and best-fit SEDs of two similar galaxies are shown in the top left panel.  These are de-redshifted (top right), and scaled to match (bottom left), effectively doubling the resolution of the photometry. With a significant number of galaxies, a composite SED with impressive spectral resolution ($R\sim50$ in the near-UV to optical) can be derived from photometric observations alone. An example is shown in the bottom right, with photometric observations in gray and median points in purple.}
	\label{fig:method}
	\end{figure*}


The construction of composite SEDs requires grouping galaxies together based on SED shape, as determined from multi-wavelength photometry.
This method is based on the work presented in \citet{Kriek2011}, with minor changes made in \citet{Forrest2016} and \citet{Forrest2017a}.

We begin by selecting a sample over some redshift range, based on Easy and Accurate $z_{phot}$ from Yale \citep[EAZY;][]{Brammer2008} outputs included in the \zfourge\ catalogs \citep{Straatman2016}.
EAZY fits linear combinations of sets of input galaxy spectral templates to photometry allowing calculation of photometric redshifts and rest-frame colors.
Combined with the medium-bands of \zfourge, this yields precise photometric redshifts, which are necessary to minimize scatter in the resulting composite SEDs.

The strength of the composite SED method is only realized when different redshifts are used.
Grouping galaxies over a narrow redshift range does not improve sampling of the rest-frame wavelengths over observations of an individual galaxy.
Therefore it is important that the redshift range of galaxies being considered is broad enough to enable continuous spectral coverage via deredshifted photometry.
\citet{Kriek2011} used a redshift range of $0.5<z<2.0$, \citet{Forrest2016} required $1.0<z<3.0$ and \citet{Forrest2017a} was based on composite SEDs from galaxies in the range $2.5<z<4.0$.
The overlap in redshift ranges was to increase the sample size in the \citet{Forrest2017a} work.
We regenerate composite SEDs from the latter two redshift ranges using the publicly released set of \zfourge\ catalogs.

The signal to noise cut for our selection is $SNR_{K_s}>20$.
In general this limits the galaxies in the sample to those which have well-defined SEDs through accurate photometry.
Combined with the similarity index described below, this ensures that two identical galaxies with observations different due only to noise determined by our $SNR$ cut will be grouped together.
Finally, we eliminate stars and other contaminants by requiring the catalog flag \texttt{use=1}, and remove X-ray selected, IR selected, and radio selected active galactic nuclei (AGN) hosts as identified in \citet{Cowley2016}.
These cuts produce 7351 galaxies in $1<z<3$ and 1294 galaxies in $2.5<z<4$.

\subsection{Grouping Method} \label{Grouping}

Once we have our sample, we run each galaxy through EAZY, using nine templates from \citet{Fioc1999, Brammer2008, Erb2010, Whitaker2011} shown in Figure \ref{fig:EAZY}.
These templates and the \Ks\ luminosity prior used are described in Section 5.1 of \citet{Straatman2016}.
Using these best fits, we generate synthetic photometric points in 22 rest-frame filters for every galaxy.
These rest-frame filters have their center points at wavelengths
$\log_{10}(\lambda_{c,i}/\textrm{\AA}) = 3.13 +0.073i$, are symmetric around those points in $\log_{10}$ space, equivalent in width in $\log_{10}$ space, and have responses of unity between their bounds.
Thus, they weight every wavelength of those between $1226 < \lambda/\textrm{\AA} < 49580$ equally in $\log_{10}$ space.

Between any two galaxies, we only compare those filters which lie between the rest-frame wavelengths photometrically observed for both galaxies.
Thus, galaxies at vastly different redshifts will have fewer filters compared-- this is taken into account when choosing a sample redshift range.
The rest-frame synthetic photometry,  $f_\lambda^{rf}$, is used to obtain a metric describing the similarity of any two galaxies as in \citet{Kriek2011}:
\begin{align}
b_{12} &= \sqrt{\frac{\Sigma(f_{\lambda}^{rf1}-a_{12}f_{\lambda}^{rf2})^2}{\Sigma(f_{\lambda}^{rf1})^2}}\\
a_{12} &= \frac{\Sigma f_{\lambda}^{rf1}f_{\lambda}^{rf2}}{\Sigma (f_{\lambda}^{rf2}) ^2}
\end{align}
Here, $b$ measures the difference between the shapes of two galaxies' SED fits, while $a$ is a scaling factor to account for flux differences.
If two galaxies have $b<0.05$, we consider them to be \textit{analogs}.

After calculating this $b$-parameter for combinations of all galaxies that passed our cuts, we look for the galaxy with the largest number of analogs, which we term the \textit{primary}.
We then take the primary and its analogs out of our list of galaxies and set them aside.
This process is repeated until the primary galaxy has fewer than 5 analogs.
Some of the analog galaxies selected due to similarity to an early primary may in fact be more similar to a primary selected later in the process.
Each analog galaxy is therefore compared to all the primaries and reassigned to the group whose primary is most similar (smallest $b$-value).
This finalizes the grouping method for the composite SEDs.

In what follows we work only with groups of at least 19 galaxies (with two exceptions), which allows for good characterization of the intrinsic SED shapes (see Sections \ref{CSED_24} \& \ref{CSED_13}).
Groups of galaxies that passed our cuts but were not placed into composite SEDs due to their small group numbers were inspected as well-- these are susceptible to noisy observations.
While we require $SNR>20$ for the \Ks\ detection bandpass, other bands for these galaxies may have lower $SNR$.
If photometry in several bands is particularly noisy in the same direction, a group of galaxies may fail the similarity criteria and be placed into separate groups.

As a result, many of these small groups look very similar to other composites in e.g., the optical wavelengths, but offset with noisy observations in the e.g., near-infrared.
While the possibility exists that these are an intrinsically separate population, these galaxies are a larger fraction of the $2.5<z<4$ sample consistent with the effects of noise.
Regardless, no group appears to have a drastically different SED shape overall, and merging a group with another similar SED shape would not effect our results due to their small numbers.

The associated observed photometry for each galaxy in a composite SED is deredshifted using \zfourge\ redshifts and scaled using the $a$ value from Equation 2, which in concert probe the underlying SED with greater resolution than is possible with photometry of a single galaxy alone.
We split these deredshifted, scaled photometric points into rest-frame wavelength bins with equal numbers of observations.
The bins therefore are not necessarily equal in wavelength width, nor are they the same between different composite SEDs.
Medians of the de-redshifted, scaled photometry in each wavelength bin are taken, generating the composite SED, as shown in Figure \ref{fig:method}.

There are non-detections in the data, particularly for quiescent galaxies in the UV, and we include these when calculating the composite SED points (i.e., negative fluxes are included when calculating medians).
If the median signal for the analog points in a bin has $SNR<1$, the associated composite SED point is considered an upper limit. 
This is often seen in the far-UV and near-IR regions of the composite SEDs where there is little flux relative to instrument sensitivities.
The final sets of composite SEDs are shown in the Appendix.

\subsection{Custom Composite SED Filter Curves}

Median values of the de-redshifted, scaled photometric values in each wavelength bin are the composite SED points.
Each of these median points also has an associated composite filter response curve, which is a linear combination of the de-redshifted photometric filters.
A given filter curve is compressed into the observed galaxy rest-frame and scaled (using a value $k$) such that there is equal area ($C$) under the resulting response curve:
\begin{eqnarray}
\lambda_{comp} &=& \lambda_{filter, rest} / (1+z) \\
C &=& k\int_{\lambda_{comp, min}}^{\lambda_{comp, max}}  R_{filter, rest}d\lambda_{comp}
\end{eqnarray}
These deredshifted, scaled filter curves are then summed to obtain the composite SED filter curve.
This method ensures that each photometric observation is equally weighted and contributes the same amount to the composite filter response curve.
The filter curves allow the characterization of the composite SEDs using EAZY and Fitting and Assessment of Synthetic Templates \citep[FAST;][]{Kriek2009}.

\subsection{Composite SEDs at $2.5<\MakeLowercase{z}<4.0$} \label{CSED_24}

The \zfourge\ catalogs have 1294 galaxies at $2.5<z<4.0$ with the requisite SNR, use flag, and non-AGN identifiers.
Of these, 944 (72.9\%) are placed into 16 groups based on SED similarity.
The resulting composite SEDs are comprised entirely of blue galaxies, which are not particularly dusty (90\% have \Av$\leq0.9$ mag).
An analysis of the sample shows that around 100 of those not initially placed in a group are in fact dusty star-forming galaxies or quiescent galaxies (based on position in the \UVJ\ diagram).
However, their SED shapes are different enough to not be grouped together using the above method.
For these populations we increase the $b$-parameter cutoff to $b<0.15$ to recover 2 \UVJ-quiescent groups (44 galaxies) and 2 dusty star-forming groups (49 galaxies), all of which show slightly more scatter than our blue composite SEDs.
In total we therefore have 20 composite SEDs comprised of 1037 galaxies (80.1\% of the sample that passed our cuts).

The 90\% mass completeness of \zfourge\ at $z=3$ is $\log_{10}(M_{90}/M_\odot)\sim10$ \citep{Tomczak2016}.
However, there are a number of galaxies with strong \OIII\ and \Hbeta\ emission in our detection bandpass, \Ks.
We therefore are sensitive to objects with particularly strong emission from these lines at lower masses than those galaxies without this emission.

The method used to generate these composite SEDs has small methodological changes to that used in \citet{Forrest2017a}.
These changes allow inclusion of a larger number of galaxies in the composite SEDs.
The Extreme and Strong Emission Line Galaxies from \citet{Forrest2017a} are now split into several composite SEDs, the differences largely driven by the UV slope of a galaxy.

\subsection{Rebuilding Composite SEDs at $1<z<3$ from \citet{Forrest2016}} \label{CSED_13}

For consistency with the new composite SEDs constructed here, we also rebuild composite SEDs at $1.0<z<3.0$ using the publicly released \zfourge\ catalogs (v3.4).
The composite SEDs presented in \citet{Forrest2016} used an earlier version of the \zfourge\ catalogs.
This version did not use the same deep stacked \Ks-band detection image, and thus was limited to 3984 galaxies in the $1<z<3$ sample which also met the other requirements above, namely having $SNR_{K_s}>20$ and \texttt{use=1}.
Using the updated catalogs, we obtain 7351 galaxies with the same criteria.
The resulting 71 composite SEDs have 6314 galaxies, or 85.9\% of the original sample and unlike the initial grouping at $2.5<z<4.0$ include a number of quiescent and dusty star-forming composite SEDs.
One of the groups with fewer than 19 galaxies is also of interest however, as it contains 14 galaxies with very blue colors and strong emission features, consistent with the emission line galaxies seen in the higher redshift sample.
We thus include this composite SED in our following analysis.
\zfourge\ completeness is $\log_{10}(M_{90}/M_\odot)\sim9$ at $z=1.5$ \citep{Tomczak2016}, and 524 (8.3\%) galaxies in our $1<z<3$ sample are less massive than this due in part to H$\alpha$ falling in the \Ks\ bandpass at $2<z<2.5$.

Between the two sets of composite SEDs, there are 6921 total galaxies, i.e., there are 444 galaxies which fall in the redshift range $2.5<z<3$ and are in composite groups in both regimes.
In this work, we use only these newly constructed composite SEDs, and do not use those previously studied in \citet{Forrest2016, Forrest2017a}.

\section{Measuring Individual Galaxy and Composite SED Properties} \label{Measure}


	\begin{figure}[tp]
	\centerline{\includegraphics[width=0.5\textwidth,trim=0in 0in 0in 0in, clip=true]{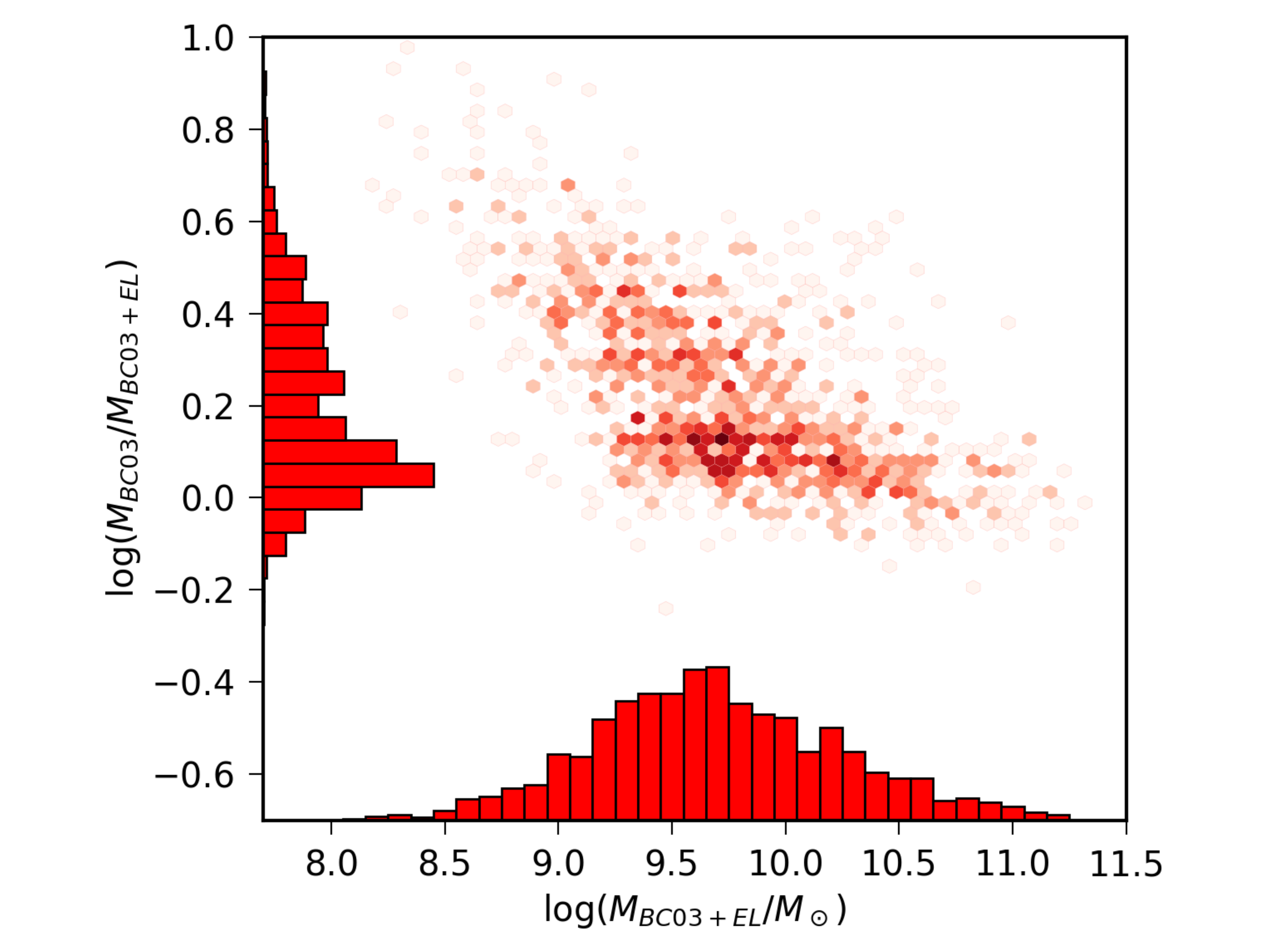}}
	\caption{Differences in best fit mass from FAST for galaxies in our $2.5<z<4$ sample.  The masses of low mass galaxies are significantly overestimated if the effects of strong emission lines are not accounted for. These emission lines show effects on galaxies up to $\log(M/M_\odot)\sim10$.}
	\label{fig:deltaFAST}
	\end{figure}


In this Section we discuss the measurement of quantities which are used in our analysis (Section \ref{SAnalysis} and Section \ref{PAnalysis}).
For our analysis of the composite SEDs, we consider both the properties of the analog galaxies and the properties of the composite SED itself.
When composite SED `fluxes' are described, these values are scaled due to the construction method of the composite SED.
As a result, these can only be used validly as part of a color.

\subsection{Rest-frame colors}

We consider the \UVJ\ diagram in our analysis.
Rest-frame fluxes for analog galaxies are taken from the \zfourge\ data release.
These values are calculated using EAZY and the nine different galaxy templates mentioned above.
We use this same method with our composite SEDs and their custom filter curves to generate rest-frame colors for each composite SED.

\subsection{Using Emission Line Templates with \textit{FAST}}

As shown in previous work, failure to account for emission lines when fitting templates to galaxy photometry can lead to severe errors in parameter estimation for the strongest emitters \citep[e.g.,][]{Stark2013, Salmon2015, Forrest2017a}.
We therefore refit all of the galaxies in our sample using FAST \citep{Kriek2011} and a series of models from \citet{Bruzual2003} (BC03) with emission lines added. 

These emission lines are based on modeling done with CLOUDY 08.00 \citep{Ferland1998}, with methods from \cite{Inoue2011} and \cite[][ see Section 3.2]{Salmon2015}.
Briefly, the ionization parameter, metallicity, and density of hydrogen are varied to produce sets of emission line ratios from Lyman-$\alpha$ to 1 $\mu$m.
These emission lines are added to the BC03 high resolution models and are used in our FAST runs.
We use a \citet{Chabrier2003} IMF, a \citet{Kriek2013a} dust law, and an exponentially declining star formation history.
All of these assumptions can have effects on our results, in particular dust and age determinations.
We do not explore these issues in depth here, but refer the reader to \citet{Cassara2016a, Leja2017} for more information.

We refit all galaxies in our composite SED samples with this set of emission line models, allowing other parameters to range as in the \zfourge\ catalogs.
No galaxies were assigned an age greater than the age of the universe at the corresponding photometric redshift.
The differences from these new fits and the \zfourge\ results are non-negligible, showing two main groups (see Figure \ref{fig:deltaFAST}).

The first group consists of galaxies with emission lines, for which models sans emission lines overestimate the mass by $0.75\pm0.12$ dex at $\log_{10}(M/M_\odot)\sim8.5$, decreasing to agreement at $\log_{10}(M/M_\odot)\sim10.5$.
The second group does not have strong emission features, and the masses are therefore consistent between the two fits.
On average, this second group is higher mass, and the greater stellar continua reduce the effects of any nebular emission lines on SED fitting, although some galaxies down to $\log_{10}(M/M_\odot)\lesssim9.5$ show little evidence of emission.

The composite SEDs are also fit with FAST.
Similar to fluxes, the output masses and SFRs are scaled to unphysical values, although properties such as sSFR, age, and dust attenuation ($A_V$) are unaffected.
For such affected properties, we use the median of the analog population as a characteristic value for the composite SEDs.

\subsection{UV Slope}

We fit a power law to the composite SED points within the wavelength range $1500<\lambda/$\AA$<2600$, $F\propto\lambda^\beta$ to obtain the UV slope, $\beta$.
This effectively prevents contamination from Lyman-$\alpha$ emission, as changing the Lyman-$\alpha$ template flux yields no change in the fit UV slope.
We also masked around the 2175\AA\ dust feature and refit the power law.
For the vast majority of composite SEDs this makes no difference to the fit.
In the several cases which show clear attenuation at this wavelength, we mask points over $2000<\lambda/$\AA$<2350$ and use the resultant exponent.

\subsection{\Dfour}


	\begin{figure}[tp]
	\centerline{\includegraphics[width=0.45\textwidth,trim=0in 0in 0in 0in, clip=true]{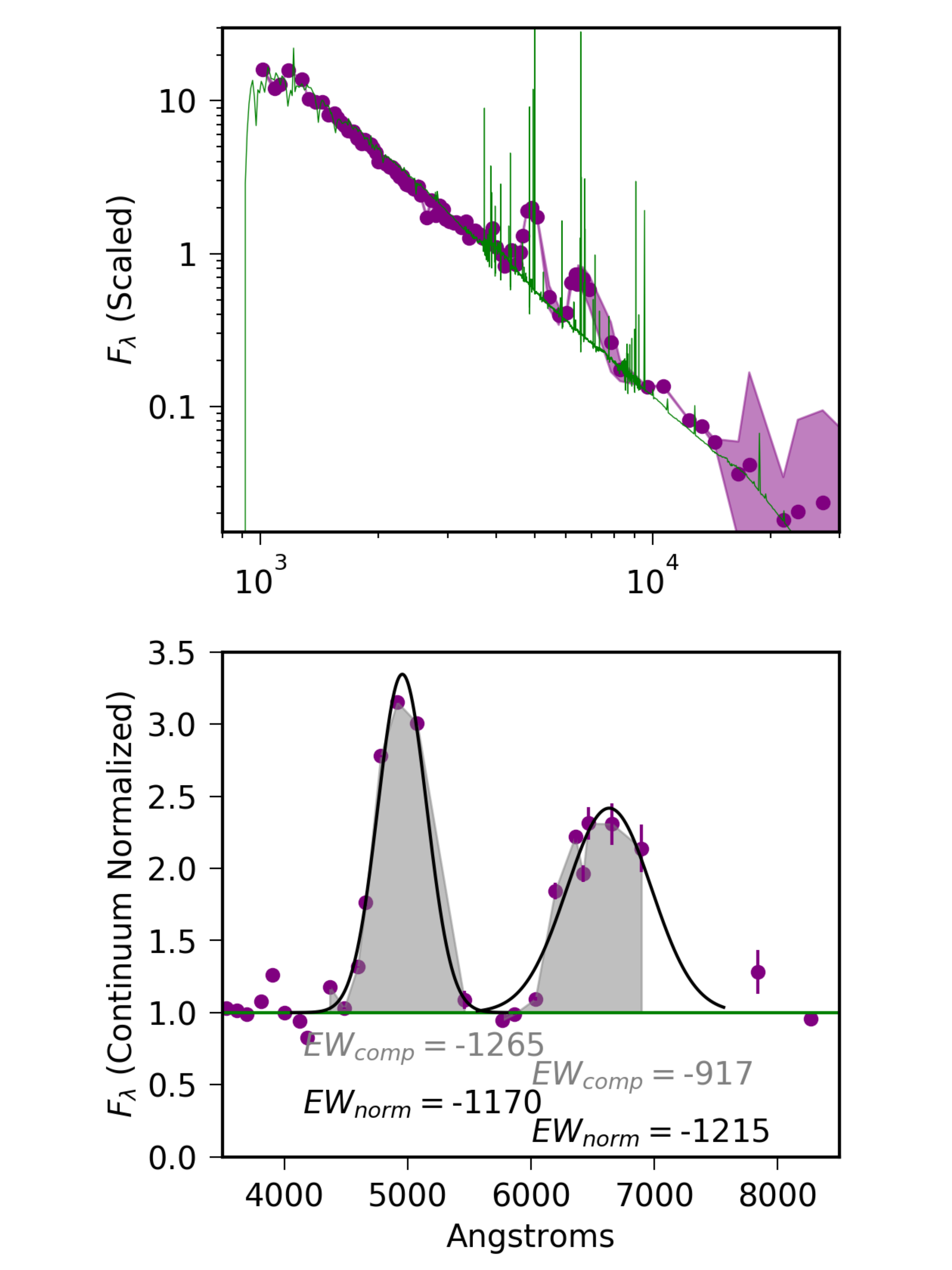}}
	\caption{Equivalent width determination. \textit{Top:} The bluest composite SED from galaxies at $1<z<3$ as determined by UV slope, $\beta$. The median composite SED points and associated errors on medians are shown in purple. The best fit emission line SED is in green. \textit{Bottom:} The continuum normalized flux of the composite SED points showing \OIII+\Hbeta\ and H$\alpha$ emission.  The black curves show fits of Gaussian profiles to the emission line blends, while the gray shading shows a simple trapezoidal integration to obtain the equivalent width. In general these two methods agree within 10\%, although in cases of extreme emission such as this, the selection of points for trapezoidal integration is an important factor and can lead to larger discrepancies. Throughout this work, we quote equivalent widths from the Gaussian curve fits.}
	\label{fig:EW}
	\end{figure}


The 4000 \AA ngstrom break (\Dfour) is defined in \citet{Bruzual1983} as
\begin{eqnarray}
\Dfour = \frac{ (\lambda_{blue}^2-\lambda_{blue}^1) \int_{\lambda_{red}^1}^{\lambda_{red}^2} f_\nu d\lambda}{ (\lambda_{red}^2-\lambda_{red}^1) \int_{\lambda_{blue}^1}^{\lambda_{blue}^2} f_\nu d\lambda},
\end{eqnarray}
with $(\lambda_{blue}^1, \lambda_{blue}^2, \lambda_{red}^1, \lambda_{red}^2) = (3750, 3950, 4050, 4250)$ \AA.
Given the limited resolution of our composite SEDs, these integrals generally correspond to two points on either side of the break, but are still well constrained.

Several of the ELG composite SEDs have \Dfour$<1$.
This indicates stellar populations dominated by light from young, massive O stars \citep[e.g.,][]{Poggianti1997},  and is also influenced by any nebular continuum emission that is present \citep{Byler2017,Byler2018}.
Our composite SED band width also means that our \Dfour\ calculation is sensitive to the Balmer break and strong emission from [OII]$\lambda 3727$, which for the most extreme emitters could lower our measured \Dfour\ by up to 0.2.
Errors are determined by calculating \Dfour\ using the 1$\sigma$ error flux values for the composite SED points.
As detailed in Appendix C of \citet{Kriek2011}, our photometric redshift errors are sufficiently small such that they will not effect this measurement.

\subsection{Equivalent Widths}

We measure the rest-frame equivalent width of [OIII]$\lambda$5007,4959+H$\beta\lambda$4861 for all of our composite SEDs and H$\alpha$+[NII]+[SII] for our $1<z<3$ composite SEDs.
For the $2.5<z<4.0$ sample, the H$\alpha$+[NII]+[SII] line blend falls between the $K_s$-band and the \textit{IRAC} 3.6 $\mu$m filter, and will therefore not be observable until the \textit{James Webb Space Telescope} (\textit{JWST}) is taking data.
To measure the equivalent widths of these line blends, we use the best fit SEDs from FAST models with emission lines, as described above.
We remove the emission lines from these best-fit SEDs to obtain the stellar continuum, and convolve this with the custom composite SED filters to obtain synthetic photometry of the continuum.
The composite SED is then normalized by this synthetic photometry.
 
Several ways of measuring the equivalent width were tested, two of which are shown in Figure \ref{fig:EW}.
First, we perform a simple trapezoidal integration under the continuum normalized composite SED in the area of interest
\begin{eqnarray}
EW_{\textrm{[OIII] } blend} = \int_{4361}^{5507} (1-f_\lambda/f_c) d\lambda\\
EW_{H\alpha\textrm{ }blend} = \int_{5763}^{7363} (1-f_\lambda/f_c) d\lambda,
\end{eqnarray}
where $f_\lambda$ is the composite SED flux and $f_c$ is the continuum flux from the best fit SED.
We note that the composite SED points themselves must be within these limits and therefore are nominally in a narrower wavelength regime.
However, since the custom composite SED filters are fairly broad, signals outside of these wavelength limits are in fact being probed.
This would be the case even if a single composite SED point were used.

In addition, we fit a Gaussian profile to the continuum normalized composite SED and integrate under that curve.
The results are generally similar to within 10\%.
However in some cases, the composite SED points have spacing which yields a discrepancy between the two methods, as can be seen with the H$\alpha$ emission in Figure \ref{fig:EW}.
In these cases, the fits were visually inspected, and in all such cases the Gaussian profile fit was judged to be superior.

For blends of multiple lines, such as [OIII]$\lambda$5007,4959 + \Hbeta$\lambda$4861, we also attempted fitting multiple Gaussian curves, one to each line.
Forcing the center of each Gaussian profile to be at the emission wavelength provides a good overall fit to the data, but the individual curves are often unphysical, usually showing strong absorption in one Gaussian profile and strong emission in another.
Further constraining this multi-Gaussian profile fit by forcing a line ratio, e.g., [OIII]$\lambda$5007/[OIII]$\lambda$4959=3, generally results in fitting absorption for \Hbeta, which we take to be unphysical as well given the large H$\alpha$ EWs.
The overall fits are again good, and very similar to the fit of the single Gaussian curve above.
Equivalent widths measured from the Gaussian profiles are in both cases within a few percent of the single curve fit.
The broadness of the custom composite SED filters is the cause of this, as we do not accurately resolve out the different lines.

Weak emission is difficult to quantify accurately, especially when the continuum fit is not good or the composite SED is noisy relative to the line.
In general, we are confident in emission equivalent widths down to 20\AA, and most composite SEDs have \OIII+\Hbeta\ and H$\alpha$ equivalent widths greater than this.
In the remainder of this paper, referenced equivalent widths will be from the single Gaussian profile fit for each line blend, and all values will be in the rest-frame.

\subsection{Morphology}

The \zfourge\ data release includes a catalog of sources cross-matched with the \textit{CANDELS} morphological catalogs of \citet{VanderWel2012}.
The resolution of the $HST-F160W$ imagery used in these catalogs is 0.06" after drizzling.
While at high redshifts this nominally makes fitting small galaxies difficult, \citet{VanderWel2012} find that galaxies with half-light radii of 0.3 pixels are recovered correctly using {\sc galfit} \citep{Peng2010}.
There are 31 galaxies in our sample across a range of redshifts and composite SEDs that have fit sizes below this limit-- excluding these galaxies makes no difference in our results.
We compare sizes and S\'ersic indices for galaxies of different classifications in Section \ref{morph}.


	\begin{figure*}[tp]
	\centerline{\includegraphics[width=0.9\textwidth,trim=0in 0in 0in 0in, clip=true]{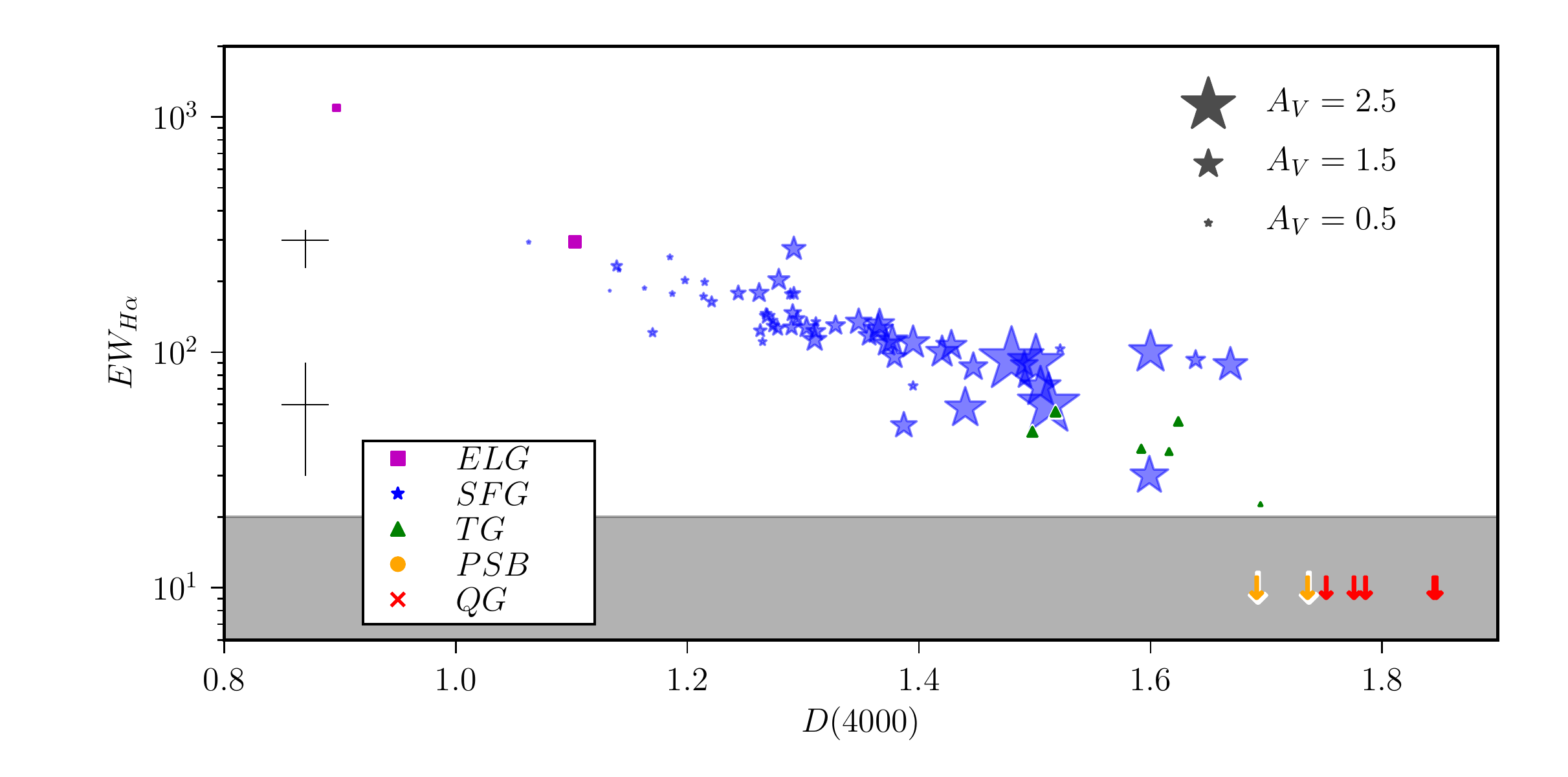}}
	\caption{H$\alpha$ $EW_{\rm REST}$ against \Dfour\ for our $1<z<3$ composite SEDs.  This is used in concert with the \OIII+\Hbeta\ $EW_{REST}$ and dust attenuation fit using FAST, indicated by marker size, to classify the composite SEDs which show evidence of star formation. Star Forming Galaxy composite SEDs are blue stars, showing a trend toward larger dust attenuation and lower H$\alpha$ $EW_{\rm REST}$ at higher \Dfour. Transition Galaxies (green triangles) show significantly less dust for their \Dfour, bucking the trend of the other star-forming galaxies.  Those classified as Extreme Emission Line Galaxies are shown as magenta squares, which have \Dfour$<1.1$ as well as $EW_{[OIII]}>400$\AA. Post-Starburst Galaxies (orange), and Quiescent Galaxies (red) are not detected above our noise threshold of 20\AA\ (gray shaded region). Representative error bars are shown on the left. We emphasize that these are errors on the composite SED measurements and do not convey the scatter in the underlying galaxy populations.}
	\label{fig:spec}
	\end{figure*}



	\begin{figure}[tp]
	\centerline{\includegraphics[width=0.5\textwidth,trim=0in 0in 0in 0in, clip=true]{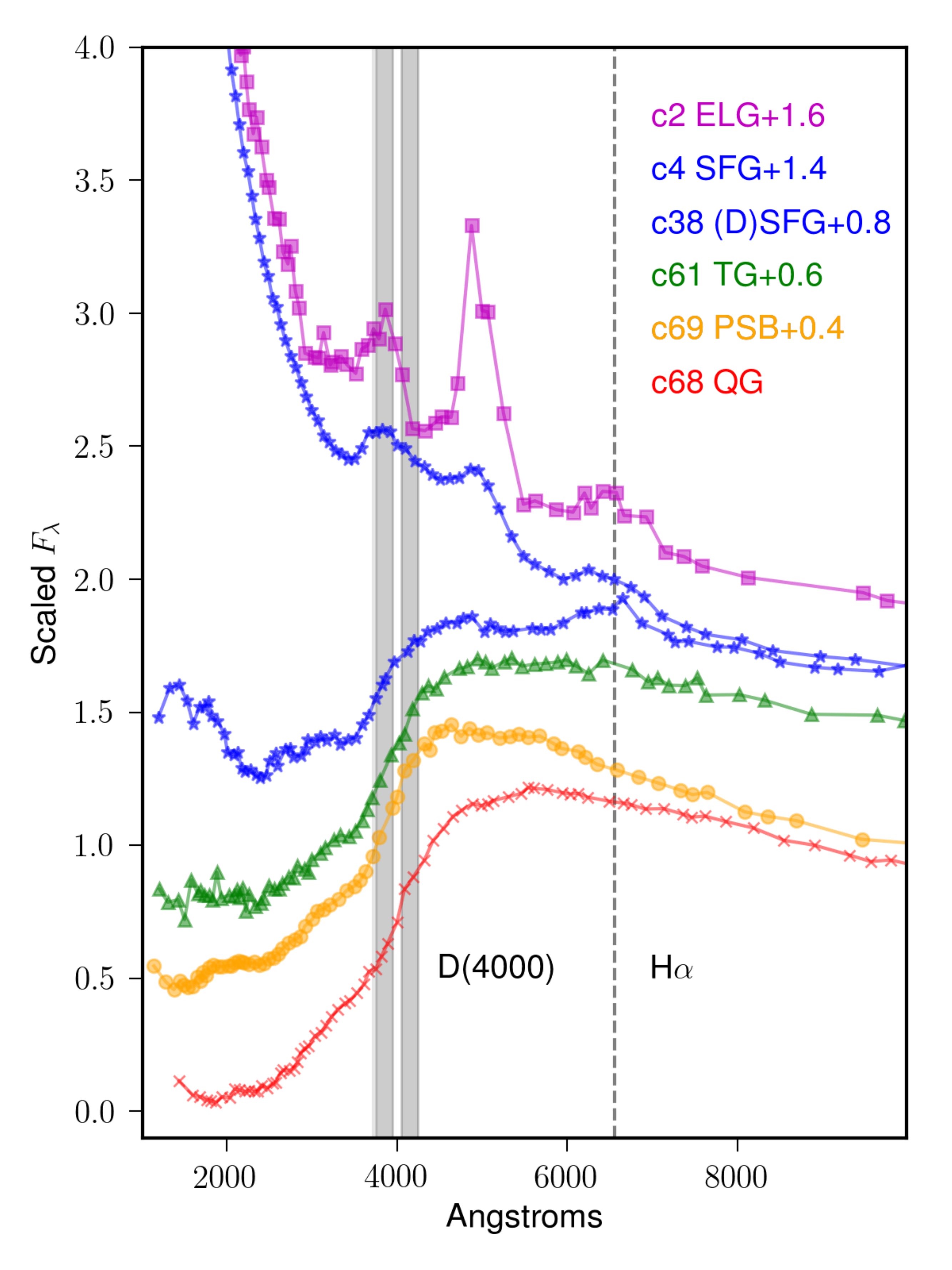}}
	\caption{Several representative composite SEDs showing the rest-frame optical wavelengths.  These are plotted with a vertical offset for clarity. The composite id for reference with the Appendix is given. We are able to discern between the quiescent and post-starburst composite SEDs due to the sharper turnover of the post-starbursts.  That is, the spectral peak redward of the 4000\AA\ break is blueward of $\sim4500$\AA\ for post-starbursts, while older quiescent populations peak redward of 5000\AA. Blue star-forming galaxies and extreme emission line galaxies have considerably more UV-optical flux than any of the other types shown here. Id's given are for reference with data in the Appendix - all composite SED's shown here are from the $1<z<3$ set.}
	\label{fig:opt}
	\end{figure}
	

\section{Spectral Feature Analysis} \label{SAnalysis}

\subsection{Composite SED Classification}

In this work we classify our composite SEDs which show evidence of star formation by their \Dfour, H$\alpha$ and \OIII\ emission line strengths, and dust attenuation (see Figure \ref{fig:spec}).
\Dfour\ is a proxy for age \citep[e.g.,][]{Poggianti1997}, although with a dependence on metallicity \citep[e.g.,][]{Kauffmann2003a}.
H$\alpha$ probes the star formation activity for galaxies in a composite SED \citep[e.g.,][]{Kennicutt2012}.
While \OIII\ emission is dependent upon abundances, it is also sensitive to ionizing photons from young stars.

 It should be noted that both emission features as measured from the composite SEDs are blends.
H$\alpha$ is blended with [NII] and [SII] lines, but will dominate the signal for strongly star-forming galaxies; while \OIII\ is blended with \Hbeta, the oxygen will similarly dominate for the strongest emitters \citep[e.g.,][]{Baldwin1981, Kewley2013}.
Using these parameters derived from the composite SEDs means that this selection is independent of the morphologies of the galaxies involved, and is less sensitive to photometric errors than color selections for individual galaxies.
Nonetheless as described below, we still pick out trends in both parameters based on our classification.

The majority of our composite SEDs have equivalent widths of $EW_{H\alpha}\sim100$\AA\, and these are classified as Star Forming Galaxies (SFGs).
With increasing D(4000) we see this EW decrease, as well as an increase in dust attenuation as fit by FAST, in agreement with Figure 8 from \citet{Kriek2011}.
However, there are several composite SEDs with \Dfour$\gtrsim1.5$ and $30\lesssim EW_{H\alpha}$/\AA$\lesssim50$ which show less dust than other composite SEDs at similar values.
These are classified as Transition Galaxies (TGs), which will be discussed in greater detail in Section \ref{Disc}.

At low \Dfour\ we see groups with large $EW_{H\alpha}$ (and $EW_{[OIII]+H\beta}>400$\AA), which we classify as Extreme Emission Line Galaxies (ELGs).
A slightly different set of composite SEDs with many of the same galaxies is discussed in more detail in \citep{Forrest2017a}.

While the SFGs have $D{\rm (4000)}\sim1.3\pm0.2$ and $\log_{10}(EW_{H\alpha}/$\AA$) \sim2^{+0.2}_{-0.1}$, several composite SEDs have \Dfour$>1.5$ and $EW_{H\alpha}<20$\AA.
Upon visual inspection, we classify these as either Quiescent Galaxies (QGs) or Post-Starburst Galaxies (PSBs) based on the sharpness and location of the turnover of the SED around 5000\AA.
While dusty SFGs, TGs, and QGs all have a plateau in the SED from $0.5-0.7\mu$m (in $F_\lambda$ units), the PSBs have a distinct peak blueward of this, consistent with the populations of A-type stars that helped lead to their original moniker-- E+A galaxies.
Figure \ref{fig:opt} shows the optical wavelengths for examples of the different classes.

The composite SEDs constructed from galaxies at $2.5<z<4$ lack coverage across wavelengths to which H$\alpha$ is redshifted-- the line falls between the \Ks-band and the \textit{IRAC} channels.
We again use \Dfour\ and $EW_{[OIII]+H\beta}$ to identify 3 ELG composite SEDs, and use visual identification to compare the others to the low redshift sample.
There is less variety seen than at $1<z<3$, with 15 of the 19 composite SEDs clearly falling into the star-forming regime, including the two dusty composite SEDs.
The remaining two, constructed from \UVJ-quiescent galaxies, show some scatter, but appear most similar to the PSBs from the $1<z<3$ sample.
While there may be a few older quiescent galaxies in these samples, they are in the minority.

In what follows, we compare the properties of galaxies in these different classes.
On the whole, reassigning a single composite SED to a different class (within reason, i.e., SFG to/from TG or PSB to/from QG) does not affect our conclusions.
Throughout the paper, we will use purple to represent ELGs, blue for SFGs, green for TGs, orange for PSBs, and red for QGs.

\subsection{$EW$-mass}

The use of deep narrowband imaging to find emission line galaxies in specific redshift windows has been used for over two decades \citep[e.g.,][]{Hu1996, Cowie1998, Teplitz1999}, notably in the High Redshift Emission Line Survey \citep[HiZELS; ][]{Geach2008}.
More recently, emission line galaxies have also been identified from flux excesses in broadband filters relative to adjacent multi-wavelength photometry \citep[e.g.,][]{Fumagalli2012, Finkelstein2013, Labbe2013, Stark2013, Smit2014}.
Composite SEDs have been used for emission line galaxy selection as well \citep{Kriek2011, Forrest2017a}.

Using these large numbers of equivalent widths, trends have been found with mass and redshift.
\citet{Fumagalli2012} use data from \textit{3D-HST} to quantify H$\alpha$+[NII] EW against mass across several redshifts and find that for galaxies of a given mass, EWs are higher at higher redshift, similar to results from HiZELS \citep{Sobral2013}.
Similarly, data from HiZELS \citep{Khostovan2016} and \textit{Spitzer} \citep{Smit2015} have been used to trace out [OIII]+H$\beta$ EWs against mass, with similar conclusions.
Specifically, [OIII]+H$\beta$ EW for galaxies of a given mass appear to have decreased since $z\sim2.5$.

\begin{figure*}[t]
	\begin{center}
	\includegraphics[width=0.8\textwidth]{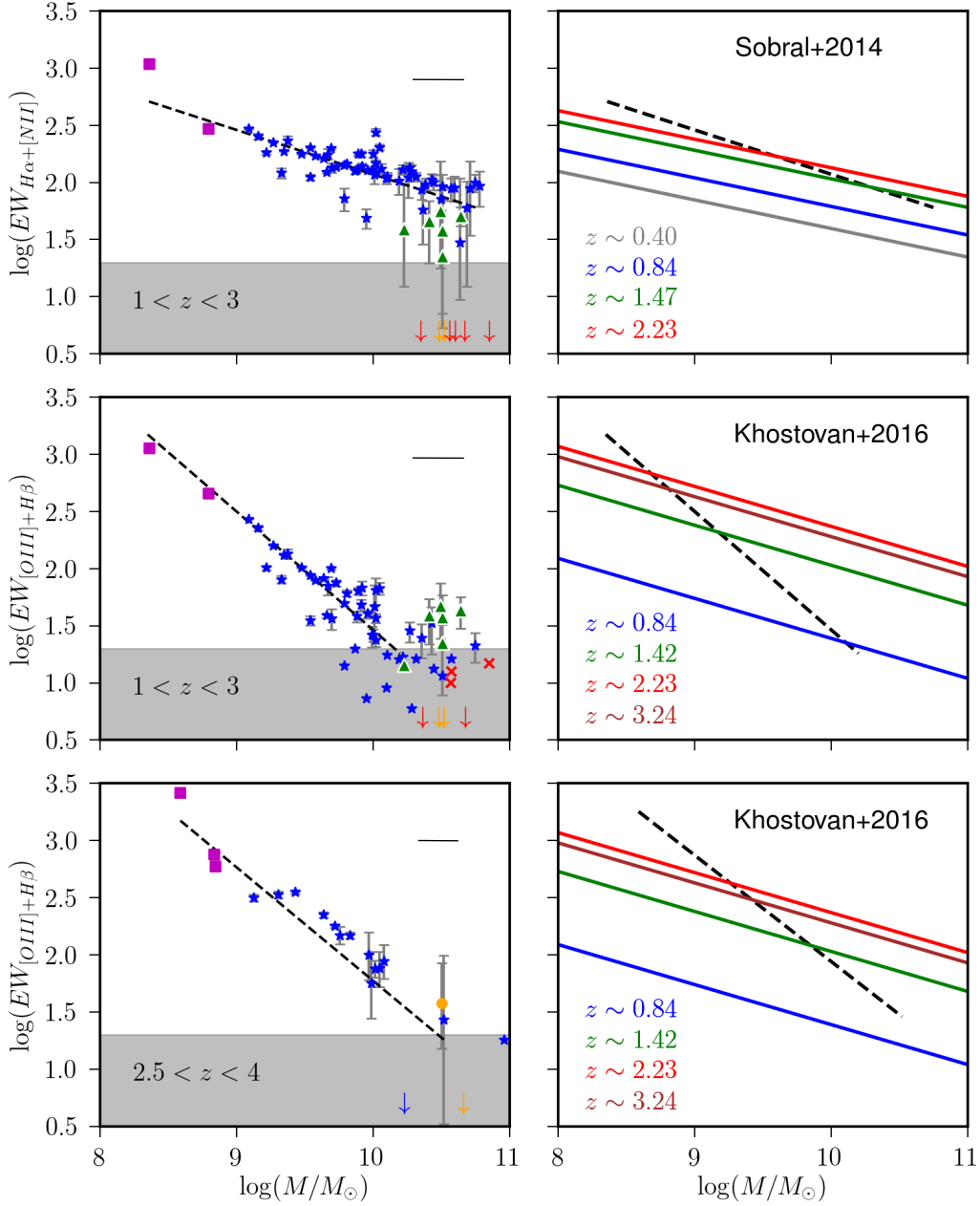}
    \caption{Equivalent widths against mass for the composite SEDs, with points colored according to the classification as in previous figures (see Figure \ref{fig:spec}). The gray shaded regions represent EW$<20$\AA, which we take to be the limit of our sensitivity with the composite SEDs. Masses are medians of the analogs in a composite SED. Typical standard deviations for the masses in a composite SED are 0.3 dex for $1<z<3$ and 0.25 dex for $2.5<z<4$, shown by the black error bar in the upper right of the left panels. \textit{Top Left}: H$\alpha$+[NII] EW for composite SEDs at $1<z<3$.  \textit{Middle Left}: \OIII+\Hbeta\ EW for composite SEDs at $1<z<3$. \textit{Bottom Left}: \OIII+\Hbeta\ EW for composite SEDs at $2.5<z<4$.  \textit{Right}: Fits to composite SED EWs shown as black lines, with relations from \citep{Khostovan2016} (top, middle) and \citep{Sobral2013} shown as colored lines.}
	\label{fig:ewo3m}
	\end{center}
	\end{figure*}

The H$\alpha$ EWs for the $1<z<3$ sample are in good agreement with both \citet{Fumagalli2012} and \citet{Sobral2013} (see top right panel of Figure \ref{fig:ewo3m}).
Unfortunately we are unable to probe H$\alpha$+[NII] in our $2.5<z<4$ sample to see if this ratio varies with redshift, but this will be explored by \textit{JWST}.

Interestingly, our results for \OIII+\Hbeta\ diverge from HiZELS work \citep{Khostovan2016}.
In the $1<z<3$ sample we have good agreement at $\log(M/M_\odot)\sim9$, but more extreme emitters and fewer massive emitters.
The picture is similar in $2.5<z<4$ except that the samples agree at $\log(M/M_\odot)\sim9.5$.

We note that our sample is not mass-complete down to the lowest masses, as only low mass galaxies with strong emission lines in the \Ks-band will be included.
As seen in Table \ref{TC}, the composite SEDs do not have any galaxies of similar mass to the ELGs (below $\log(M/M_\odot)\sim9$) without such remarkable emission.
As a result, our large $EW$ (low mass) end of the sample is skewed upward.
Also, the composite SEDs are not sensitive to weak emission that can be found in more massive star forming galaxies.
\citet{Khostovan2016} note these factors in the HiZELS sample as well, but find that these biases do not effect the $EW-$mass relation significantly.

The remaining difference between our samples is the width of our redshift bins, across which lines move in and out of the \Ks-band (our detection bandpass).
At $2<z<2.5$, H$\alpha$ falls into the \Ks-band and \OIII+\Hbeta\ does the same at $3<z<3.8$.

Regardless, the TGs clearly show reduced H$\alpha$ emission relative to SFGs of the same mass.
Combined with their elevated \OIII+\Hbeta, this suggests the possibility of AGN.
While the strongest AGN should be removed with the catalogs from \citet{Cowley2016}, the possibility of low level AGN contamination does remain.
Rest frame optical spectroscopic follow-up will allow quantification of such contamination.


\section{Photometric Analysis} \label{PAnalysis}

\subsection{Color Relations}

\begin{figure*}[tp]
	\centerline{\includegraphics[width=0.7\textwidth,trim=0in 0in 0in 0in, clip=true]{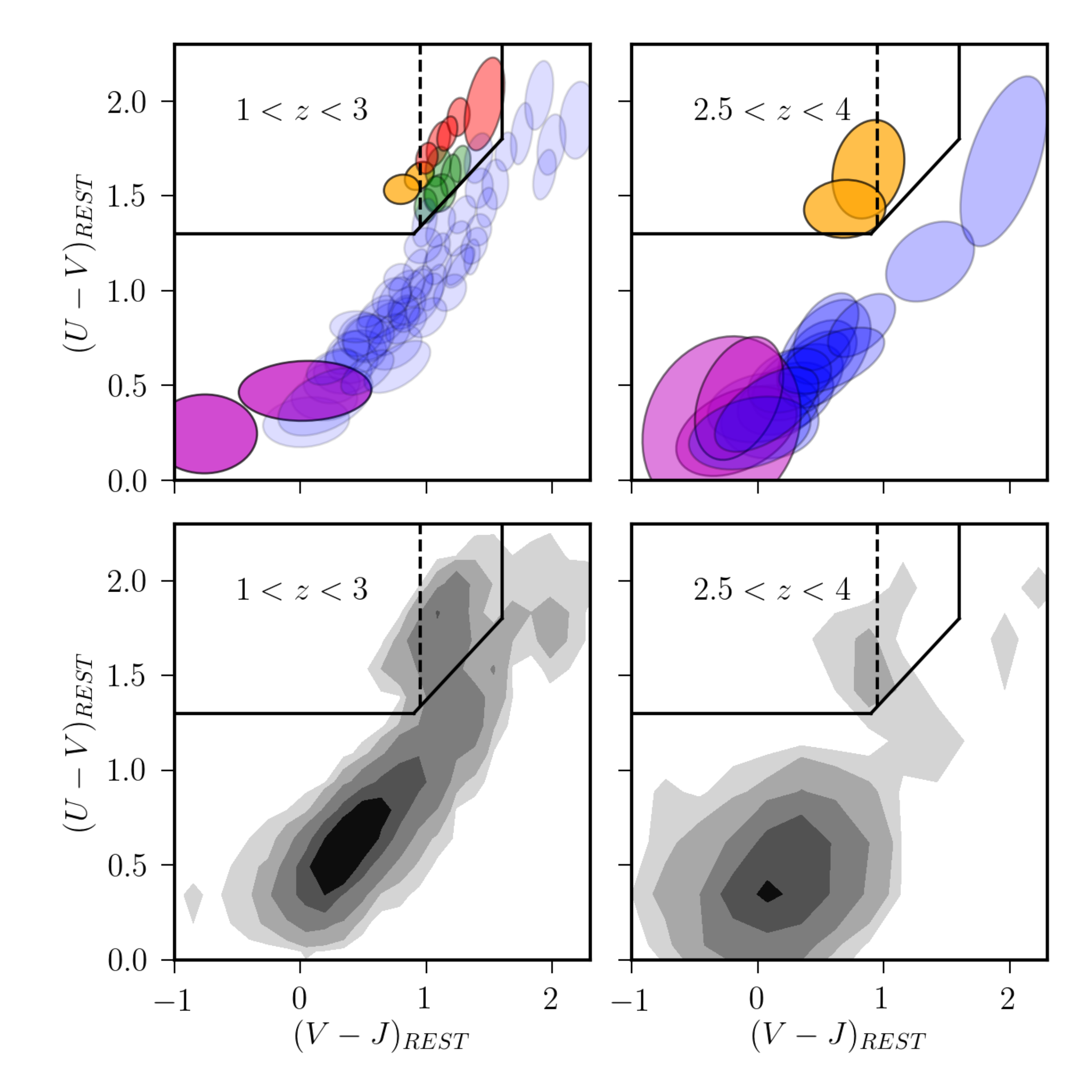}}
	\caption{\UVJ\ diagram. \textit{Top:} The $1\sigma$ error ellipses of our composite SEDs based on analog positions on the \UVJ\ diagram. $1<z<3$ composite SEDs are on the left, while $2.5<z<4$ composite SEDs are on the right.  Star-forming composite SEDs are shown in blue, emission line galaxies in magenta, post-starbursts in orange, quiescent composite SEDs in red, and transitional composite SEDs in green. The vertical dashed line is from \citet{Whitaker2012b, Wild2014} and separates post-starbursts (blueward) from older quiescent galaxies (redward). \textit{Bottom:} Contours of analog galaxies on the \UVJ\ diagram. Contours for the $1<z<3$ sample are 3, 10, 30, 100, and 300 galaxies, while $2.5<z<4$ contours are 3, 8, 22, 60, and 120 galaxies. }
	\label{fig:uvj}
	\end{figure*}

The composite SEDs are formed based on multi-color comparisons.
As such, we would expect the groups to separate into distinct groups on color-color diagrams, the best known of which is the \UVJ\ diagram \citep[e.g.][]{Wuyts2007,Williams2009,Whitaker2012,Straatman2016,Forrest2016}.
There is a spread in the colors of analogs in a given composite SED, and we display these by calculating $1\sigma$ error ellipses based on the covariance between the colors, shown in Figure \ref{fig:uvj}.
As expected, composite SEDs in a given class are mostly separated from other classes, although some of the individual galaxy colors do overlap.
This indicates that while the \UVJ\ diagram does a good job on average discerning between a simple red and blue sequence, it does not yield the whole picture that can be obtained by analyzing the full SED of a galaxy.
In this picture, the colors of TGs are consistent with galaxies in the green valley and with the transition galaxies of \citet{Pandya2017}.


We note that there is reduced diversity in the $2.5<z<4$ composite SEDs.
While some of this is due to the reduced sensitivity to objects with faint stellar continua, this does not explain the lack of quiescent objects, nor the lack of transition objects, as \zfourge\ is mass complete for these samples out to $z\sim3.5$.
This is suggestive that these populations are rarer at high redshifts, which is known to be the case for quiescent and dusty objects \citep[e.g.,][]{Spitler2014, Straatman2016, Glazebrook2017}.
Nonetheless, post-starburst galaxies are found here, implying that star formation has been turned off, or at least significantly reduced, as studies have shown that galaxies in this regime of the \UVJ\ diagram can still be forming stars, albeit with low sSFR \citep[e.g.,][]{Ciesla2017}.

Additionally, the star-forming sequence of the \UVJ\ diagram broadens, suggesting a wider range of colors for star forming galaxies at high redshift.
While measurement errors may play a small role here, the intrinsic spread is expected to increase due to the presumed bursty nature of star formation in young galaxies \citep[e.g.,][]{Papovich2001, Castellano2014, Izotov2016}, although uncertainties remain on this front \citep[see, for example,][]{Smit2015}.
There are also a greater number of galaxies with strong nebular emission falling in the rest-frame V band, which boosts galaxies to particularly blue colors in \VJc.

\begin{figure*}[tp]
	\centerline{\includegraphics[width=0.7\textwidth,trim=0in 0in 0in 0in, clip=true]{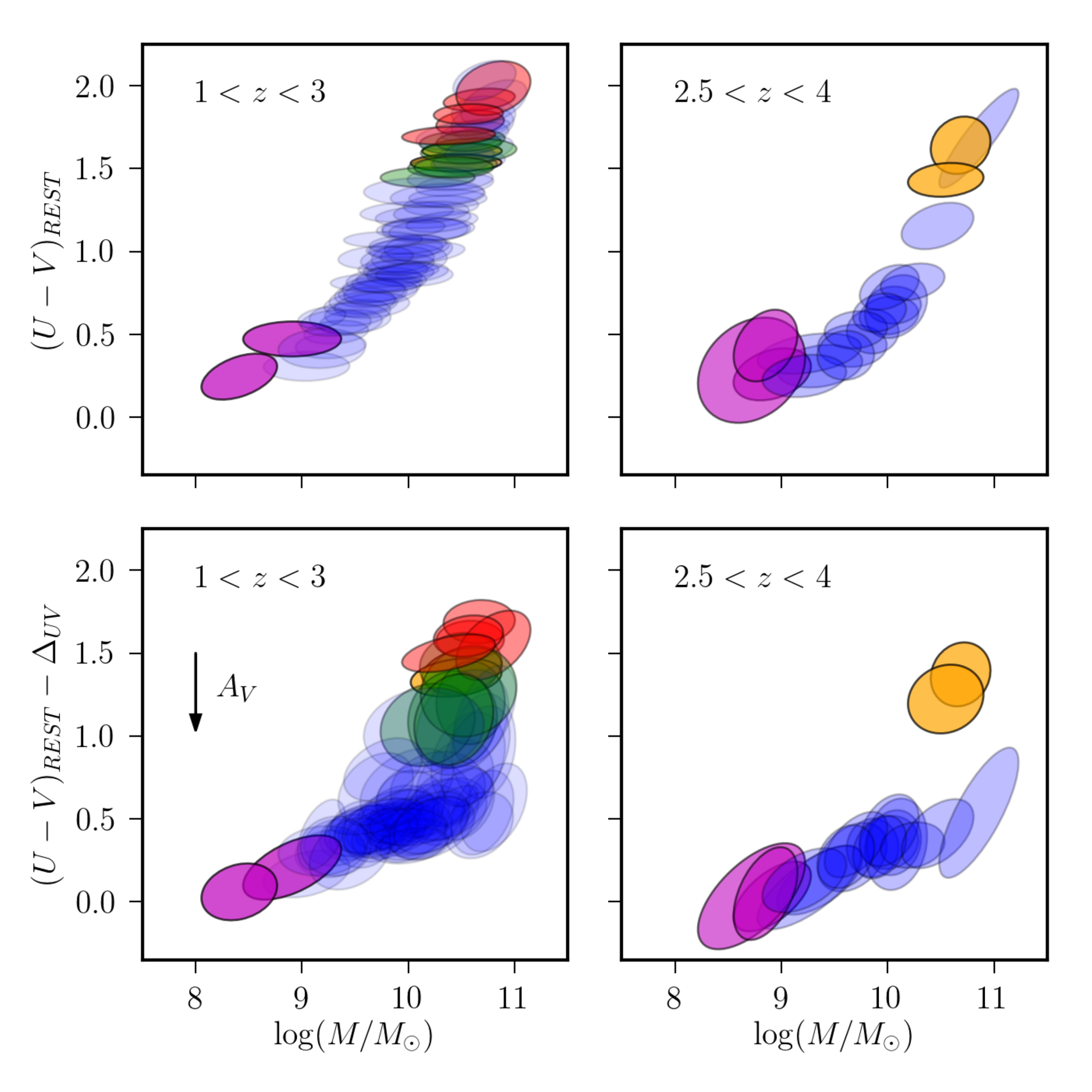}}
	\caption{Composite SEDs plotted as $1\sigma$ error ellipses of the analogs that comprise that composite SED in \UVc-mass. The top rows are colors fit using EAZY, while the bottom row is corrected by dust attenuation derived using FAST.  The left column is for galaxies in our sample in $1<z<3$, while the right column is for galaxies in $2.5<z<4$.  The dust correction removes many of the star-forming galaxies in the observed green valley.}
	\label{fig:UVm_comp}
	\end{figure*}

This classification scheme is also consistent with that determined using a color-mass diagram.
We correct the rest-frame \UVc\ colors using the dust attenuation for a galaxy as described in \citet{Brammer2009} and shown in Figure \ref{fig:UVm_comp}.
This correction for dust attenuation more closely approximates the intrinsic colors, providing a clearer separation between dusty SFG, TG, PSB, and QG composite SEDs.


\subsection{Star Forming Main Sequence}


	\begin{figure}[tp]
	\centerline{\includegraphics[width=0.5\textwidth,trim=0in 0in 0in 0in, clip=true]{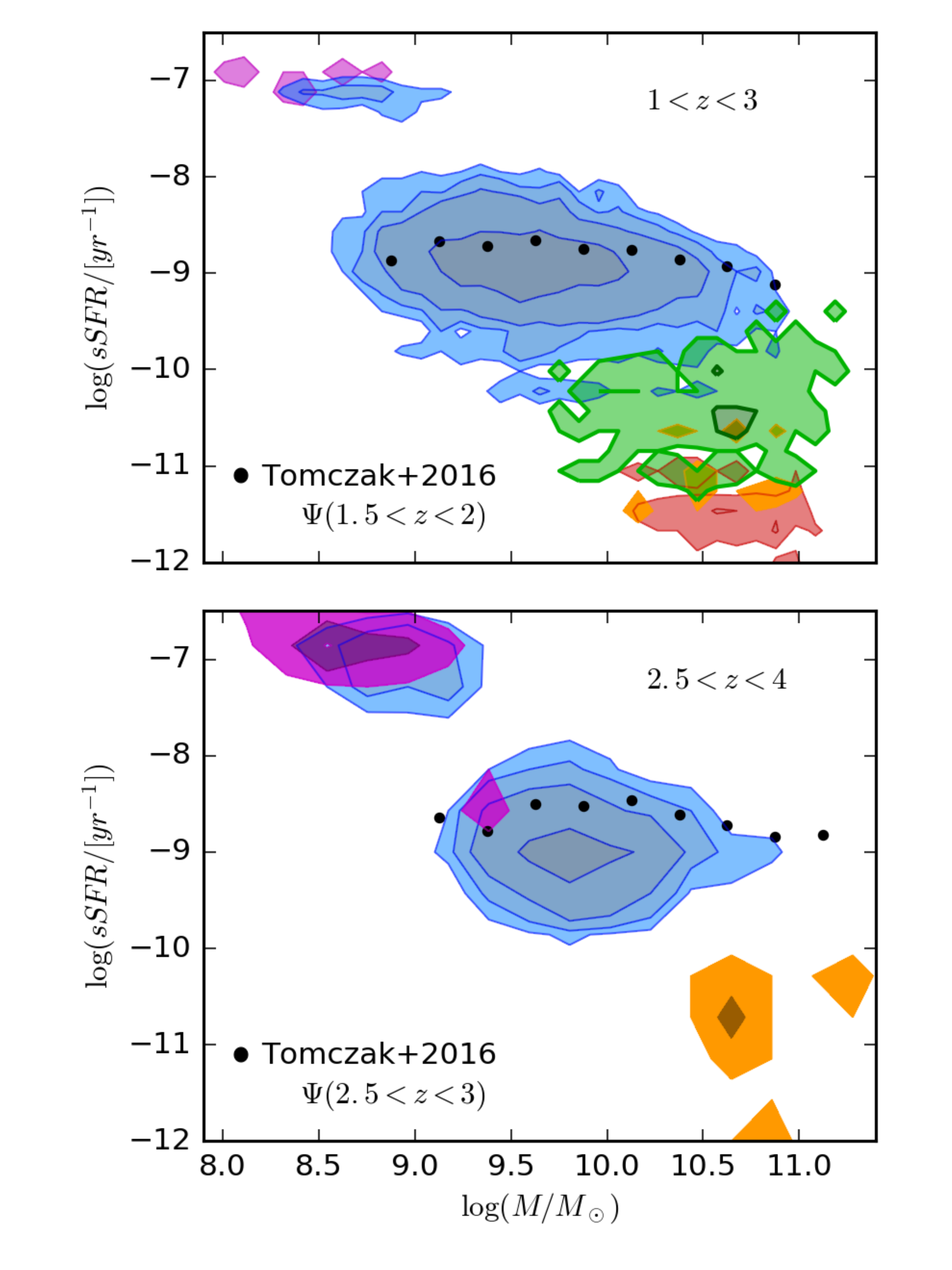}}
	\caption{Specific star formation rate-stellar mass relation for galaxies in different composite SEDs classes.  The contours show all galaxies in composite SEDs of a specific classification, using the same color scheme as previous figures.  The various classes show separation with respect to sSFR.  The black points are the $SFR-M_*$ relations for star-forming galaxies from \citet{Tomczak2016} at similar redshifts.  While there is some overlap between the TGs and PSBs, the mean sSFR for PSBs is lower.  The average sSFR increases at higher redshifts.}
	\label{fig:sfms}
	\end{figure}


Previous works have also classified galaxies in narrow redshift bins based solely upon sSFR \citep[e.g.,][]{Pandya2017}.
Figure \ref{fig:sfms} shows the locations of individual galaxies of different composite SED class on the sSFR-$M_\odot$ plane.
While on the whole different classes do separate out nicely, there exists some overlap between TGs and PSBs in the $1<z<3$ redshift set.
The sSFRs are lower for the PSBs on average, which is reasonable since they are thought to be almost completely quenched, while TGs are in the process of quenching.
However, numerous studies have shown an evolution of SFR (and sSFR) against mass as a function of redshift-- in general, higher redshifts  show fewer quenched galaxies, higher mass galaxies quenching, and higher star formation rates for star-forming galaxies of a given mass \citep[e.g.][]{Whitaker2012, Behroozi2013, Sparre2015, Tomczak2016, Pandya2017}.
Due to the large width of the redshift bins for our composite SEDs, the evolution of these relations is a driver of the scatter observed in Figure \ref{fig:sfms}.
Therefore while we find larger numbers of galaxies with high sSFRs and fewer quenched galaxies at higher redshifts, we do not make any conclusions about the efficacy of galaxy categorization by sSFR.

\subsection{Morphological Evolution \label{morph}}

\begin{figure*}[tp]
	\centerline{\includegraphics[width=0.7\textwidth,trim=0in 0in 0in 0in, clip=true]{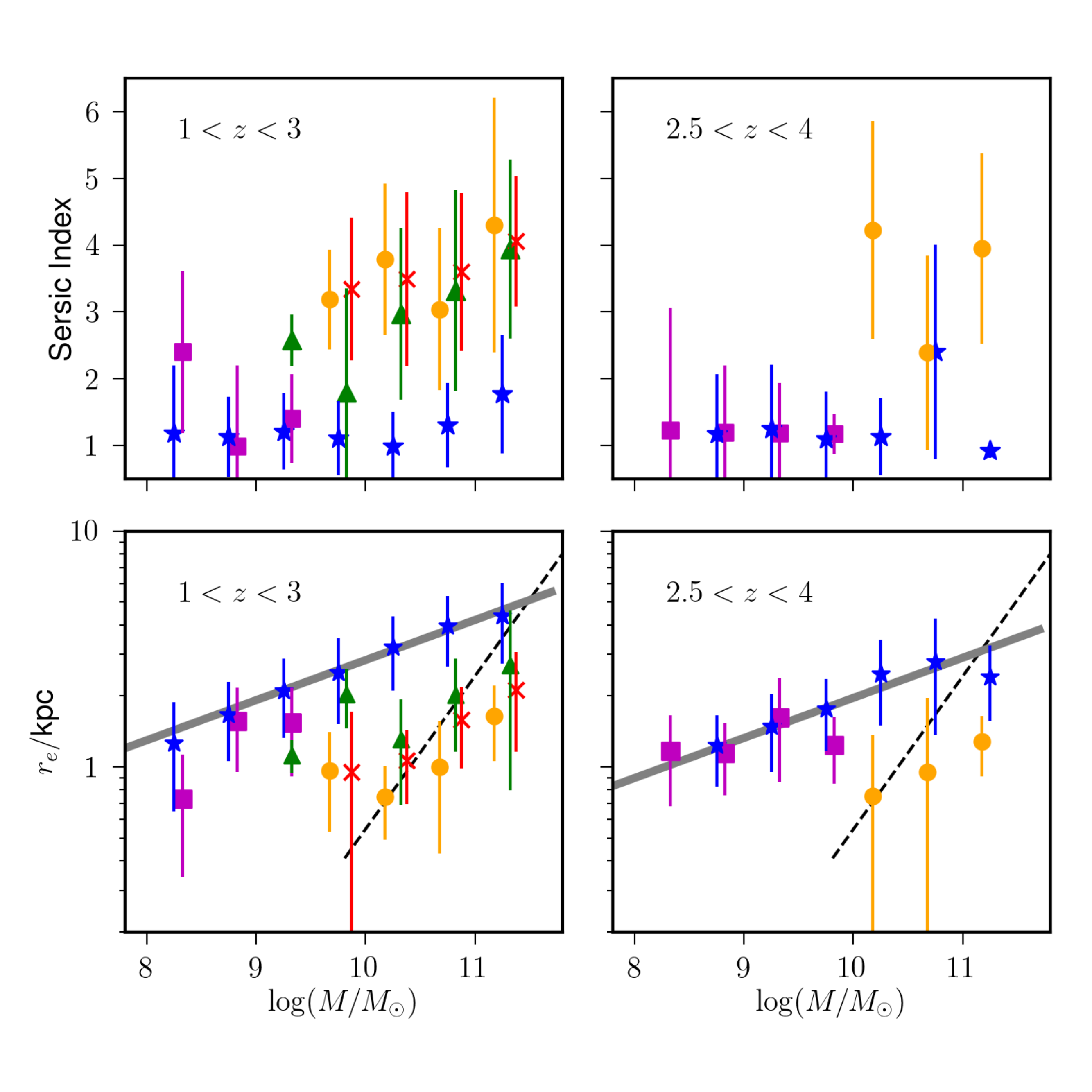}}
	\caption{Morphological characteristics of galaxies in different composite SED classes. \textit{Top:} The S\'ersic indices for galaxies in our sample according to mass and classification, color coded as in previous figures.  Points are slightly offset along the abscissa for clarity  and error bars show the 16-84\% range in values for analog galaxies in composite SEDs of the class and binned mass range.  \textit{Bottom:} The size-mass plane for galaxies in our composite SEDs. The SF galaxies follow the size-mass relations from \citet{Allen2017} (thick gray line) quite well, while at low-redshift all other classes are smaller in size for a given mass (left).  At $2.5<z<4$ (right), the ELGs have similar sizes, while PSBs are smaller.  In both cases, the non-star-forming classes lie near the compactness selection criterion of \citep{Barro2013}, shown as a dashed line.}
	\label{fig:size_mass}
	\end{figure*}

We also investigate the morphologies of galaxies with regard to mass and classification, shown in Figure \ref{fig:size_mass}.
The star-forming galaxies match well with previous analyses of the size-mass relation \citep[e.g., \zfourge\ and COSMOS/UltraVISTA;][] {Allen2017, Faisst2017}.
Additionally, most of the TGs, PSBs, and QGs lie near the selection
criterion for compact quiescent galaxies from \citep{Barro2013}.

At $1<z<3$, the SFGs have larger sizes than all other galaxy classifications for a given mass.
The TGs in particular, have median sizes half those of the SFGs and twice those of the QGs, as would be expected for galaxies whose star formation is being quenched.
Meanwhile, the sizes for PSBs are on average smaller than QGs of the same mass ($\log(p)\sim-7$ from a Kolmogorov-Smirnov, or K-S, test).

ELGs and SFGs have similar S\'ersic indices of $n\sim1$, and the PSBs and QGs have values of $n\sim3.5$ (the distributions are quite similar, with $p=0.68$ from a K-S test, in agreement with results from \citet{Almaini2017}).
Meanwhile the TGs have values of $n\sim2-3$.
Combined with the H$\alpha$ EWs, this indicates that morphological changes such as the development of a central bulge are already underway before star formation has ceased completely, although further size growth may occur \citep[see also ][]{Papovich2015}.

Additionally, we fit lines to the S\'ersic indices of the analog galaxies in a given class against both mass and redshift.
No class shows evidence for significant evolution with redshift, with slopes $\mid \Delta n/\Delta z \mid<0.2$, smaller than the spread and errors on the values.
All classes except the ELGs show median increases with mass, although such increases are $\Delta n<1$ over $8.5<\log(M/M_\odot)<11.5$, no larger than the distribution of galaxy values for a given mass.


\subsection{Post-Starburst and Transitional Galaxy Number Densities}


	\begin{figure}[tp]
	\centerline{\includegraphics[width=0.5\textwidth,trim=0in 0in 0in 0in, clip=true]{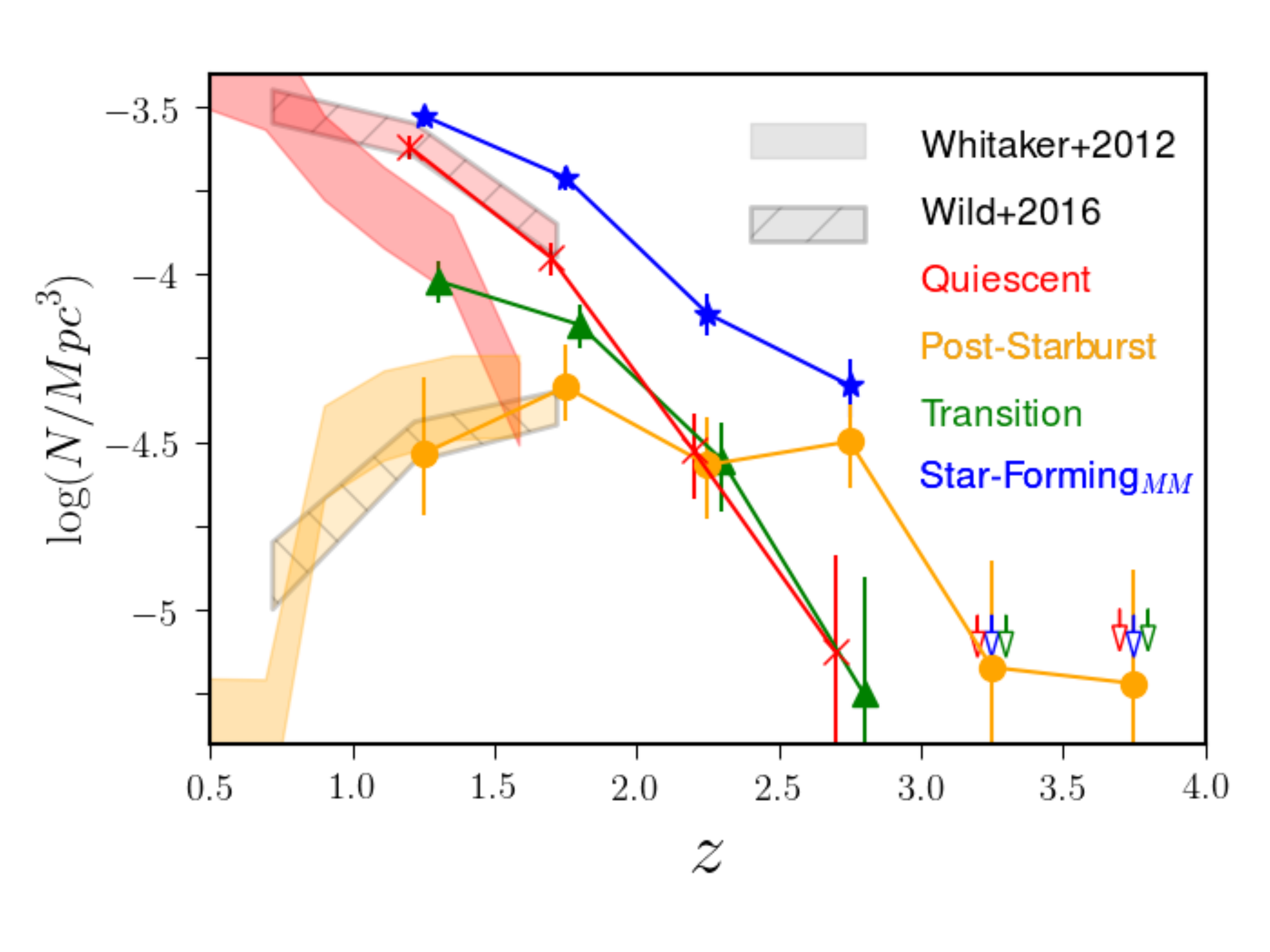}}
	\caption{Comoving number densities of QGs (red), PSBs (orange), TGs (green), and mass-matched SFGs (blue) against redshift.  Our results are consistent with the results from \citet{Wild2016} shown as hashed shaded regions.  Results from NMBS \citep{Whitaker2012b} are shown as non-hashed shaded regions.  Notably, the shapes of the TG and QG curves appear quite similar, which is suggestive of them being along a similar evolutionary pathway. While both these tracks flatten out towards lower redshifts, the PSBs show strong evidence for a turnover around $z\sim1.5$.}
	\label{fig:psbfrac}
	\end{figure}


We calculate the number densities of TGs, PSBs, and QGs of galaxies in our composite SEDs across redshift space.
Additionally, we show number densities for a mass-matched population of massive star-forming galaxies, achieved by selecting composite SEDs above a median mass of $\log(M/M_\odot)>10.25$ and including all galaxies in those composite SEDs.
The mass limit was chosen by maximizing the $p$-value from a two sample K-S test between the masses of the TGs and selected SFGs ($p=0.69$).
Incidentally, this also yields $p=0.62$ for masses of the PSBs and selected SFGs.
Since a galaxy's stellar mass shouldn't change significantly during the quenching process (ignoring mergers), these mass-matched SFGs  should be most similar to progenitors of the TGs and PSBs. 

Our results for the PSBs and QGs, shown in Figure \ref{fig:psbfrac}, are consistent with those from the Newfirm Medium Band Survey \citep[NMBS;][]{Whitaker2012b} and the UKIDSS Deep Survey \citep[UDS;][]{Wild2016} at $z\sim1-2$  and extend out to higher redshifts.  The $z\sim1$ side also lines up with results from the Galaxy and Mass Assembly and VIMOS Public Extragalactic Redshift Surveys \citep{Rowlands2018}.
We note that each of these works selects PSBs in a different manner-- \cite{Whitaker2012b} use an age motivated cut on the \UVJ\ diagram and \cite{Wild2016} use a selection based on PCA colors, while we use composite SEDs to select on population \Dfour\ and emission line characteristics.

Comparing the number densities of different groups across a range of redshifts suggests that the transitional phase is even rarer than the traditional post-starburst phase at high redshifts, but becomes more common at $z<2$.
Additionally, the density of PSBs is relatively constant from $1.5<z<3$, with evidence for a turnover at $z\lesssim1.5$, below which such galaxies become rarer.
While the PSB curve stays mostly flat, the shape of the TG curve is more similar to that of the QGs, which increases dramatically from $z=3$ before beginning to flatten at $z\sim1.5$.
This suggests that the TG population represents a quenching mechanism with a longer timescale than PSBs, which has become more prevalent at later times, discussed in more detail in the following section.

Across $3<z<4$, \citep{Tomczak2016} report a \zfourge\ mass completeness limit of $\log_{10}(M/M_\odot)=10.25$, in agreement with our mass matching selection.
The  TGs, PSBs, and QGs have mass distributions with medians $\log_{10}(M/M_\odot)=10.51$, 10.54, and 10.61, respectively, in close agreement to the mass-matched SFG population, with a median of $\log_{10}(M/M_\odot)=10.48$.
Due to the similar masses and detection-band magnitudes for members of the TGs, PSBs, and QGs, any biases and selection effects would effect them in a similar manner.
While some individual galaxies in our $2.5<z<4$ PSB composite SEDs could be quiescent or transitioning, the clear differences in composite SED shape guarantee that they would be few in number.
The average properties of these different classes, including the mass-matched SFG sample, are shown in Table \ref{TC}.

\section{Discussion} \label{Disc}

Our TG classification appears successful in picking out galaxies transitioning between more typical star-forming galaxies and quiescent galaxies.
These galaxies have masses $\log_{10}(M/M_\odot)\sim10.5$, which are similar to dusty SFGs, PSBs, and QGs.
However, there is no evidence of large amounts of dust in the TGs (A$_{\rm V}\sim0.7$ mag) compared to dusty SFGs with similar masses (\Av$\sim1.7$ mag), and they show less H$\alpha$ emission-- $EW_{\rm REST}\sim40$\AA\ (vs. $\sim100$\AA\ for dusty SFGs; Figure \ref{fig:ewo3m}, Table \ref{TC}).
The red colors and low emission line equivalent widths are therefore due to fewer O and B type stars and low level residual star formation rather than to heavy dust obscuration, as expected for SFGs of similar mass.

The TGs still show more dust than PSBs and QGs (A$_{\rm V}\sim0.4$ mag) and are morphologically different ($r_e$/kpc$\sim2$ vs. 1 for PSBs and $n=2.9$ vs. 3.5; Figure \ref{fig:size_mass}, Table \ref{TC}). 
K-S Anderson-Darling, and Mann-Whitney tests for the distributions of TGs and PSBs  in dust, size, and S\'ersic index reject the hypothesis that the two groups are drawn from the same distribution ($p$-values of 0.018, 0.014, 0.025 in the three tests for the S\'ersic index, and $\log(p)<-4$ for the \Av\ and size comparisons in all three tests).

The intermediate changes in morphology that occur in a galaxy while its star formation is being shut off are unclear.
Galaxies will generally be disky at early times when they are actively forming stars, and develop a spheroidal bulge which dominates the morphology at late times after star formation has ceased.
A number of recent works have proposed the idea of compaction \citep{Dekel2014, Barro2013} and morphological quenching \citep{Martig2009}, in which the process of developing this central bulge is in fact the cause of (or due to the same cause as) star formation cessation.
Unless morphological changes occur on timescales less than $\sim10$ Myr (the sensitivity of H$\alpha$ to star formation), we argue that such changes begin before star formation has been completely switched off. 

Cessation of star formation in a disk with continued star formation in a central bulge could explain the morphological and $EW$ trends seen.
 Such a process would lead to the galaxy's light being concentrated in the center yielding measurements of smaller sizes and larger S\'ersic indices while also showing H$\alpha$ emission.  The opposite process, where star formation continues in the disk but shuts off in the bulge, would not show these same effects, contradicting the observations. We do not argue against this happening for individual galaxies, but it appears to not be the case for the majority.

Galaxies in the green valley with similar low level sSFRs have had several potential explanations proposed.
The most common is that these galaxies are in the process of quenching by some as yet undetermined mechanism(s), which are likely dependent on both galaxy mass and environment \citep[see Introductions of e.g.,][for nice summaries]{Darvish2016, Kawinwanichakij2017a, Papovich2018}.
The variety of quenching mechanisms are associated with different timescales for the cessation of star formation.
\citet{Barro2013} and \citet{Schawinski2014} showed that galaxies in the green valley of the color-mass diagram are representative of multiple quenching mechanisms and not a single separate population.
On the other hand simulations have claimed that a single timescale of $\lesssim2$ Gyr to cross the green valley is able to match observations \citep[e.g.,][]{Trayford2016, Dave2017, Nelson2017, Pandya2017}.

However, there is also the possibility that quiescent galaxies have had their star formation `rejuvenated' and are thus moving into the green valley from the red side as suggested in both observations \citep[e.g.,][]{Rampazzo2007, Fang2012, Darvish2016, Pandya2017} and simulations \citep[e.g.,][]{Dave2017, Nelson2017}.
Such rejuvenation is thought to be rare, and also results in only a small change in color, which cannot move a previously quenched galaxy to match the colors of galaxies in the blue cloud \citep{Dave2017, Nelson2017}.
While the PSBs have nearly constant number densities across $1.5<z<3$ before becoming rarer at lower redshifts, the number density of the TGs in Figure \ref{fig:psbfrac} closely follows that of the QGs over the same time, suggesting an evolutionary pathway.
The similar numbers also lead us to conclude that rejuvenated galaxies are not a significant fraction of our TGs, though we cannot rule them out entirely.

Another possibility is that SFGs oscillate about the star-forming main sequence, with periods of enhanced and reduced star formation on the order of 0.3 dex \citep[e.g.,][]{Tacchella2016}.
Not only do simulations suggest this is more common for lower mass galaxies \citep{Zolotov2015a}, but our TGs also extend over 1 dex below the main sequence, implying that this explanation can only contribute a small portion of the TGs observed.

Recently, \citet{Dressler2018} have noticed a population of `late bloomers', massive galaxies at $z\sim0.5$ which have formed most of their stellar mass in 2 Gyr before that epoch.
These galaxies have some broad similarities to the TGs, including \UVJ\ position, stellar mass, and declining star formation rates.
However, they also have a wider range of SED shapes and morphological properties, preventing us from concluding that they are the similar objects.
It should be noted that beyond $z\sim2.5$ it becomes difficult to {\it not} have the majority of stellar mass formed in the 2 Gyr before observation due to the age of the universe at these times.
Galaxies with such SFHs would therefore be considerably more common.

A further hypothesis is that all galaxies in the process of quenching will have a post-starburst phase, which is shorter than the overall time in the green valley and either precedes or follows it.
The relative number densities of TGs and PSBs conflict with this idea, as the number densities of PSBs are more constant over $1.5<z<3.0$, while TGs continue to increase to low redshifts, more in concert with the QGs.

\citet{Pandya2017} showed that post-starburst (fast-quenching) galaxies are more common at high redshifts relative to the transitional (slow-quenching) galaxies which dominate the quenching process below $z\sim0.7$, in qualitative agreement with \citet{Pacifici2016}.
This is as expected, since the young age of the universe at higher redshifts prohibits any long timescale quenching from completing.
Our results are consistent with this picture, where we find spatial number densities of transitional galaxies increasing sharply with decreasing redshift, while post-starbursts appear to have a turnover at $z\sim1.5$.

As a short additional calculation, we use Equation 3 from \citet{Pandya2017}, 
\begin{eqnarray}
\langle t_{\rm TG}\rangle_{z1,z2} = \langle n_{\rm TG}\rangle_{z_1,z_2} \times\bigg{(}\frac{{\rm d}n_{\rm QG}}{{\rm d}t}\bigg{)}^{-1}_{z_1,z_2}
\end{eqnarray}
to calculate average transition time for a galaxy based on number densities of transition and quiescent galaxies at varying redshifts.
We find average transition timescales of $t_{\rm TG}\sim1.63$ Gyr at $z=1.5$ and $t_{\rm TG}\sim0.95$ Gyr at $z=2.5$, in rough agreement with \citet{Pandya2017} and slightly longer than the timescale of 1.24 Gyr from $z\sim1.5$ clusters found in \citet{Foltz2018}. 
We thus conclude that the vast majority of the TGs in our sample are in fact moving from the blue, disk dominated, star-forming cloud to the red, bulge dominated, quenched sequence, possibly through multiple mechanisms with similar timescales on the order of $1-2$ Gyr.

\begin{longtable*}{l|ccrrrrrrr}
\caption{Average Parameters of Different Classes. Equivalent widths in emission are listed as positive numbers and, along with \Dfour, are measured from the composite SEDs. Other parameters are medians of analog galaxy values, with errors shown the $16^{th}$ and $84^{th}$ percentiles.}\\
 & Class & $N_{comp}$ & $\log_{10}(M/M_\odot)$ & $EW_{[OIII]}$ (\AA) & $EW_{H\alpha}$ (\AA) & \Dfour & \Av & $r_e$ & $n$  \\ \hline \hline \endhead
$1.0<z<3.0$ & ELG & 2 & $8.66^{+0.47}_{-0.40}$ & $794^{+230}_{-230}$	& $693^{+271}_{-271}$ & $1.00^{+0.07}_{-0.07}$ & $0.45^{+0.38}_{-0.29}$	& $1.1^{+1.2}_{-0.5}$  & $1.8^{+1.3}_{-1.2}$ \\
 & SFG  	& 57	& $9.59^{+0.55}_{-0.42}$ & $34^{+48}_{-30}$	& $127^{+56}_{-37}$	& $1.30^{+0.18}_{-0.10}$	& $0.60^{+0.60}_{-0.40}$ & $2.4^{+1.8}_{-1.1}$ 	& $1.2^{+1.1}_{-0.6}$ \\
 & \hspace{1cm} SFG$_{MM}$ & 18	& $10.48^{+0.36}_{-0.42}$ & $-4^{+18}_{-19}$ & $93^{+17}_{-63}$ & $1.46^{+0.14}_{-0.08}$ & $1.70^{+0.70}_{-0.60}$ & $3.4^{+2.0}_{-1.4}$ & $1.2^{+1.1}_{-0.6}$ \\
 & TG 	 	& 6 	& $10.52^{+0.34}_{-0.51}$ & $38^{+6}_{-17}$ 	& $42^{+10}_{-7}$	& $1.60^{+0.04}_{-0.08}$	& $0.70^{+0.70}_{-0.50}$ & $2.0^{+2.1}_{-1.0}$	& $2.9^{+2.0}_{-1.4}$ \\
 & PSB  	& 2 	& $10.51^{+0.35}_{-0.38}$ & $-16^{+12}_{-12}$	& $-13^{+9}_{-9}$	& $1.71^{+0.02}_{-0.02}$	& $0.40^{+0.20}_{-0.30}$ & $1.0^{+1.0}_{-0.4}$	& $3.4^{+1.7}_{-1.0}$ \\
 & QG 	 	& 5 	& $10.61^{+0.33}_{-0.39}$ & $10^{+3}_{-14}$	& $2^{+2}_{-4}$		& $1.78^{+0.06}_{-0.02}$	& $0.40^{+0.40}_{-0.20}$ & $1.5^{+1.4}_{-0.6}$	& $3.5^{+1.8}_{-1.2}$ \\
\hline
$2.5<z<4.0$ & ELG & 3 & $8.82^{+0.72}_{-0.31}$ & $755^{+1276}_{-106}$ & -- 	& $0.94^{+0.03}_{-0.02}$	& $0.60^{+0.20}_{-0.50}$	& $1.2^{+1.1}_{-0.7}$	& $1.2^{+1.5}_{-0.9}$ \\
 & SFG 		& 15	& $9.77^{+0.39}_{-0.32}$ & $100^{+193}_{-66}$ 	& --	& $1.23^{+0.06}_{-0.06}$	& $0.40^{+0.50}_{-0.30}$	& $1.8^{+1.3}_{-0.8}$	& $1.2^{+1.4}_{-0.6}$ \\
 & PSB 		& 2  	& $10.63^{+0.32}_{-0.34}$ & $8^{+20}_{-20}$	 	& -- 	& $1.66^{+0.03}_{-0.04}$	& $0.50^{+0.31}_{-0.20}$	& $1.0^{+0.9}_{-0.5}$	& $2.8^{+1.8}_{-1.3}$\\
\label{TC}
\end{longtable*}

\section{Conclusions} \label{Conc}

In this work we have categorized $\sim 7000$ galaxies from \zfourge\ based on UV to near-IR rest-frame colors and spectral feature similarities.
Building composite SEDs allowed us to leverage the large amount of multi-wavelength photometry and accurate photometric redshifts from \zfourge\ for galaxies across a broad redshift range, $1<z<4$.
These composite SEDs show a wide range of properties and independently yield expected relations based on emission line equivalent widths, sizes, masses, and number densities.
Building composite SEDs also aided in the identification of rare populations in our sample, as well as characterization of properties that are not typically available with photometry alone.

Additionally, we find evidence for galaxies with at least two quenching patterns.
Most of these transitional galaxies show H$\alpha$ emission with $EW_{\rm REST}\sim40$\AA, star formation rates $\sim1.5$ dex beneath the star-formation stellar mass relation, effective radii half that of SFGs of similar mass, S\'ersic indices of $2-3$ increasing with mass, and colors that lie on the boundary between quiescent and star forming galaxies on the \UVJ\ diagram.
The majority of these transitional galaxies have masses  $10<\log_{10}(M/M_\odot)<11$.
The other class of these galaxies is consistent with the classical `post-starburst' regime, showing small, bulge-dominated morphologies  more consistent with quiescent galaxies ($n\sim3-4$), no nebular emission, sSFRs just above the quiescent regime, but also bluer \VJc\ colors than quiescent galaxies and transitional galaxies at similar masses and redshifts.

The greater and increasing number density of the TGs at low redshifts (0.5 dex larger than PSBs at $z=1.25$) implies that this group/quenching pathway is becoming more common, while the post-starbursts are becoming rarer at $z<1.5$.
This is potentially due to a longer timescale associated with said pathway, on the order of 1.5 Gyr, a factor of $1.5-7$ times longer than the post-starburst phase is expected to last,  and which cannot have occurred before $z\sim4$.

The process that brings star-forming galaxies into the green valley creates changes in galaxy color, sSFR, size, and S\'ersic index.
The morphologies of galaxies appear to on average begin evolution toward higher S\'ersic index before star formation ceases.
Whether this morphological evolution leads directly to star formation turning off, or if there is a common cause of both changes remains unclear, but the observations are consistent with morphological quenching.

\section*{Acknowledgments}

We wish to thank the Mitchell family, particularly the late George P. Mitchell, for their continuing support of astronomy.
We also thank the Carnegie Observatories and the Las Campanas Observatory for their assistance in making the \zfourge\ survey possible. 
Thanks also to L. Alcorn, S. Faber, B. Groves, A. Harshan, and T. Mendel for insightful conversations, as well as the anonymous referee for their helpful comments.
BF and RCK would like to thank the Hagler Institute for Advanced Study at Texas A\&M.
GGK acknowledges the support of the Australian Research Council through the award of a Future Fellowship (FT140100933).
This research made use of Astropy, a community-developed core Python package for Astronomy \citep{Astropy2013}.


\appendix{}

In this Appendix, we present the composite SEDs and their associated properties.
The composite SEDs and filter curves are available for download on \href{https://github.com/b4forrest/CompositeSEDs}{Github}.
We first list the properties derived from the composite SEDs themselves, such as equivalent widths and UV slopes in Table \ref{T1} (Table \ref{T2}) for the $1<z<3$ ($2.5<z<4$) composite SEDs.
Next are the parameters derived for individual galaxies that make up a composite SED, for which the medians, 16$^{th}$ percentile, and 84$^{th}$ percentile are given.
In the case of morphological parameters, median absolute deviations are provided, to minimize errors due to resolution limits.
These are given in Table \ref{T3} (Table \ref{T4}).
Finally, plots of the composite SEDs are shown in scaled $F_\lambda$-wavelength, all labeled with a composite id number.
For most of the composite SEDs, these are shown in individual panels, colored and separated into their classes as described in the text.
The exception to this is the SFGs at $1<z<3$, which are three to a panel.
Composite SEDs are ordered by UV slope.

\begin{longtable}{ccrrrrr}
\caption{Parameters Derived From Composite SEDs at $1.0<z<3.0$.}\\
Cid & Class & $N_{gal}$ & $EW_{[OIII]+H\beta}$ (\AA) & $EW_{H\alpha+[NII]}$ (\AA) & \Dfour & $\beta$ \\ \hline \hline \endfirsthead
Cid & Class & $N_{gal}$ & $EW_{[OIII]+H\beta}$ (\AA) & $EW_{H\alpha+[NII]}$ (\AA) & \Dfour & $\beta$ \\ \hline \hline \endhead
0 & ELG 	& 14 		& $1170^{+40}_{-40}$ 	& $1215^{+88}_{-88}$ 	& $0.90^{+0.01}_{-0.01}$ 	& $-2.35^{+0.04}_{-0.04}$ \\
1 & SFG 	& 323 	 	& $265^{+11}_{-11}$ 	& $277^{+20}_{-22}$ 	& $1.06^{+0.01}_{-0.01}$ 	& $-2.00^{+0.02}_{-0.02}$ \\
2 & ELG 	& 22 		& $461^{+14}_{-14}$ 	& $282^{+28}_{-29}$ 	& $1.10^{+0.01}_{-0.01}$ 	& $-1.95^{+0.05}_{-0.05}$ \\
3 & SFG 	& 62 		& $102^{+12}_{-11}$ 	& $183^{+21}_{-22}$ 	& $1.13^{+0.03}_{-0.03}$ 	& $-1.87^{+0.03}_{-0.03}$ \\
4 & SFG 	& 577 		& $159^{+13}_{-12}$ 	& $224^{+28}_{-29}$ 	& $1.14^{+0.02}_{-0.02}$ 	& $-1.85^{+0.03}_{-0.03}$ \\
5 & SFG 	& 537 		& $132^{+14}_{-14}$ 	& $188^{+35}_{-36}$ 	& $1.16^{+0.01}_{-0.01}$ 	& $-1.75^{+0.03}_{-0.03}$ \\
6 & SFG 	& 60 		& $36^{+10}_{-10}$ 		& $111^{+25}_{-26}$ 	& $1.27^{+0.03}_{-0.03}$ 	& $-1.74^{+0.03}_{-0.03}$ \\
7 & SFG 	& 28 		& $135^{+14}_{-14}$ 	& $233^{+34}_{-35}$ 	& $1.14^{+0.06}_{-0.07}$ 	& $-1.72^{+0.03}_{-0.03}$ \\
8 & SFG 	& 51 		& $81^{+10}_{-10}$ 		& $121^{+13}_{-16}$ 	& $1.17^{+0.01}_{-0.02}$ 	& $-1.67^{+0.05}_{-0.05}$ \\
9 & SFG 	& 34 		& $227^{+23}_{-23}$ 	& $254^{+64}_{-64}$ 	& $1.19^{+0.01}_{-0.01}$ 	& $-1.67^{+0.05}_{-0.05}$ \\
10 & SFG 	& 610 		& $102^{+10}_{-10}$ 	& $178^{+15}_{-17}$ 	& $1.19^{+0.01}_{-0.02}$ 	& $-1.60^{+0.04}_{-0.04}$ \\
11 & SFG	& 419 		& $88^{+11}_{-11}$ 		& $202^{+21}_{-23}$ 	& $1.20^{+0.03}_{-0.03}$ 	& $-1.54^{+0.04}_{-0.04}$ \\
12 & SFG 	& 503 		& $79^{+12}_{-12}$ 		& $173^{+28}_{-29}$ 	& $1.21^{+0.03}_{-0.04}$ 	& $-1.53^{+0.04}_{-0.04}$ \\
13 & SFG 	& 134 		& $39^{+10}_{-10}$ 		& $124^{+15}_{-17}$ 	& $1.26^{+0.00}_{-0.00}$ 	& $-1.50^{+0.05}_{-0.05}$ \\
14 & SFG 	& 37 		& $37^{+20}_{-20}$ 		& $135^{+65}_{-65}$ 	& $1.31^{+0.01}_{-0.01}$ 	& $-1.45^{+0.07}_{-0.07}$ \\
15 & SFG 	& 61 		& $20^{+10}_{-10}$ 		& $126^{+15}_{-17}$ 	& $1.29^{+0.04}_{-0.05}$ 	& $-1.40^{+0.07}_{-0.07}$ \\
16 & SFG 	& 41 		& $24^{+15}_{-15}$ 		& $147^{+40}_{-41}$ 	& $1.29^{+0.06}_{-0.07}$ 	& $-1.40^{+0.08}_{-0.08}$ \\
17 & SFG 	& 241 		& $82^{+10}_{-10}$ 		& $164^{+13}_{-15}$ 	& $1.22^{+0.01}_{-0.01}$ 	& $-1.39^{+0.04}_{-0.04}$ \\
18 & SFG 	& 71 		& $38^{+13}_{-12}$ 		& $130^{+54}_{-55}$ 	& $1.27^{+0.04}_{-0.04}$ 	& $-1.38^{+0.06}_{-0.06}$ \\
19 & SFG 	& 31 		& $17^{+10}_{-10}$ 		& $130^{+13}_{-16}$ 	& $1.33^{+0.02}_{-0.02}$ 	& $-1.37^{+0.08}_{-0.08}$ \\
20 & SFG 	& 20 		& $14^{+12}_{-11}$ 		& $72^{+24}_{-26}$ 		& $1.40^{+0.04}_{-0.05}$ 	& $-1.37^{+0.09}_{-0.09}$ \\
21 & SFG 	& 145 		& $50^{+14}_{-13}$ 		& $142^{+39}_{-40}$ 	& $1.27^{+0.05}_{-0.06}$ 	& $-1.33^{+0.06}_{-0.06}$ \\
22 & SFG 	& 60 		& $26^{+10}_{-10}$ 		& $137^{+13}_{-16}$ 	& $1.30^{+0.01}_{-0.02}$ 	& $-1.32^{+0.07}_{-0.07}$ \\
23 & SFG 	& 132 		& $101^{+10}_{-10}$		& $199^{+36}_{-37}$ 	& $1.21^{+0.03}_{-0.04}$ 	& $-1.32^{+0.05}_{-0.05}$ \\
24 & SFG 	& 107 		& $40^{+13}_{-12}$ 		& $128^{+31}_{-32}$		& $1.29^{+0.02}_{-0.03}$ 	& $-1.30^{+0.07}_{-0.07}$ \\
25 & SFG 	& 21 		& $16^{+13}_{-13}$ 		& $103^{+35}_{-37}$ 	& $1.52^{+0.02}_{-0.02}$ 	& $-1.28^{+0.07}_{-0.07}$ \\
26 & SFG 	& 97 		& $61^{+13}_{-13}$ 		& $146^{+37}_{-38}$ 	& $1.27^{+0.02}_{-0.02}$ 	& $-1.27^{+0.05}_{-0.05}$ \\
27 & SFG 	& 53 		& $29^{+17}_{-17}$ 		& $135^{+52}_{-53}$		& $1.35^{+0.03}_{-0.03}$ 	& $-1.16^{+0.08}_{-0.08}$ \\
28 & SFG 	& 24 		& $2^{+19}_{-18}$ 		& $89^{+74}_{-75}$ 		& $1.67^{+0.05}_{-0.07}$ 	& $-1.16^{+0.41}_{-0.41}$ \\
29 & SFG 	& 20 		& $-17^{+12}_{-12}$ 	& $95^{+26}_{-27}$ 		& $1.38^{+0.03}_{-0.03}$ 	& $-1.15^{+0.10}_{-0.10}$ \\
30 & SFG 	& 107		& $37^{+13}_{-13}$ 		& $123^{+38}_{-39}$ 	& $1.31^{+0.05}_{-0.06}$ 	& $-1.15^{+0.08}_{-0.08}$ \\
31 & SFG 	& 24 		& $46^{+17}_{-16}$ 		& $118^{+44}_{-45}$ 	& $1.36^{+0.05}_{-0.06}$ 	& $-1.13^{+0.13}_{-0.13}$ \\
32 & SFG 	& 77 		& $64^{+13}_{-13}$ 		& $178^{+33}_{-34}$ 	& $1.25^{+0.02}_{-0.02}$ 	& $-1.11^{+0.06}_{-0.06}$ \\
33 & SFG 	& 25 		& $-14^{+44}_{-44}$ 	& $110^{+182}_{-182}$ 	& $1.40^{+0.02}_{-0.02}$ 	& $-1.11^{+0.12}_{-0.12}$ \\
34 & SFG 	& 46 		& $18^{+10}_{-10}$ 		& $113^{+13}_{-15}$ 	& $1.31^{+0.02}_{-0.02}$ 	& $-1.11^{+0.07}_{-0.07}$ \\
35 & SFG 	& 38 		& $3^{+11}_{-11}$ 		& $132^{+22}_{-23}$ 	& $1.37^{+0.02}_{-0.03}$ 	& $-1.10^{+0.09}_{-0.09}$ \\
36 & SFG 	& 66 		& $75^{+14}_{-14}$ 		& $137^{+31}_{-32}$ 	& $1.27^{+0.02}_{-0.02}$ 	& $-1.10^{+0.05}_{-0.05}$ \\
37 & SFG 	& 50 		& $48^{+11}_{-10}$ 		& $178^{+15}_{-17}$ 	& $1.29^{+0.01}_{-0.01}$ 	& $-1.08^{+0.05}_{-0.05}$ \\
38 & SFG 	& 30 		& $13^{+13}_{-12}$ 		& $100^{+26}_{-27}$ 	& $1.42^{+0.03}_{-0.03}$ 	& $-0.99^{+0.10}_{-0.10}$ \\
39 & SFG 	& 26 		& $6^{+10}_{-10}$ 		& $125^{+20}_{-22}$ 	& $1.37^{+0.01}_{-0.02}$ 	& $-0.97^{+0.10}_{-0.10}$ \\
40 & TG 	& 21 		& $38^{+10}_{-10}$ 		& $38^{+13}_{-16}$ 		& $1.62^{+0.00}_{-0.00}$ 	& $-0.97^{+0.12}_{-0.12}$ \\
41 & SFG 	& 27 		& $65^{+18}_{-18}$ 		& $275^{+17}_{-19}$ 	& $1.29^{+0.06}_{-0.07}$ 	& $-0.97^{+0.10}_{-0.10}$ \\
42 & SFG 	& 36 		& $26^{+10}_{-9}$ 		& $179^{+26}_{-28}$ 	& $1.26^{+0.01}_{-0.01}$ 	& $-0.97^{+0.07}_{-0.07}$ \\
43 & SFG 	& 52 		& $69^{+10}_{-10}$ 		& $139^{+14}_{-17}$	 	& $1.30^{+0.01}_{-0.01}$ 	& $-0.96^{+0.07}_{-0.07}$ \\
44 & SFG 	& 37 		& $9^{+12}_{-12}$ 		& $109^{+31}_{-32}$ 	& $1.37^{+0.02}_{-0.03}$ 	& $-0.89^{+0.09}_{-0.09}$ \\
45 & SFG 	& 21 		& $71^{+31}_{-30}$ 		& $177^{+121}_{-122}$ 	& $1.29^{+0.03}_{-0.03}$ 	& $-0.87^{+0.09}_{-0.09}$ \\
46 & SFG 	& 50 		& $68^{+22}_{-22}$ 		& $203^{+70}_{-70}$ 	& $1.28^{+0.04}_{-0.05}$ 	& $-0.87^{+0.09}_{-0.09}$ \\
47 & SFG 	& 54 		& $16^{+17}_{-16}$ 		& $112^{+56}_{-57}$ 	& $1.38^{+0.10}_{-0.11}$ 	& $-0.86^{+0.06}_{-0.06}$ \\
48 & SFG 	& 47 		& $33^{+20}_{-20}$ 		& $106^{+69}_{-69}$ 	& $1.43^{+0.01}_{-0.02}$ 	& $-0.70^{+0.13}_{-0.13}$ \\
49 & SFG 	& 28 		& $1^{+19}_{-19}$ 		& $90^{+52}_{-53}$ 		& $1.50^{+0.00}_{-0.00}$ 	& $-0.69^{+0.23}_{-0.23}$ \\
50 & SFG 	& 26 		& $7^{+15}_{-14}$ 		& $49^{+36}_{-37}$ 		& $1.39^{+0.01}_{-0.01}$ 	& $-0.63^{+0.08}_{-0.08}$ \\
51 & SFG 	& 30 		& $25^{+12}_{-11}$ 		& $87^{+30}_{-31}$ 		& $1.45^{+0.07}_{-0.09}$ 	& $-0.55^{+0.12}_{-0.12}$ \\
52 & TG 	& 29 		& $14^{+19}_{-18}$ 		& $39^{+52}_{-52}$ 		& $1.59^{+0.05}_{-0.06}$ 	& $-0.39^{+0.11}_{-0.11}$ \\
53 & SFG 	& 35 		& $-19^{+10}_{-10}$ 	& $60^{+27}_{-28}$ 		& $1.51^{+0.03}_{-0.03}$ 	& $-0.38^{+0.26}_{-0.26}$ \\
54 & TG 	& 38 		& $39^{+11}_{-10}$ 		& $46^{+19}_{-20}$ 		& $1.50^{+0.10}_{-0.13}$ 	& $-0.38^{+0.12}_{-0.12}$ \\
55 & SFG 	& 29 		& $-11^{+17}_{-17}$ 	& $92^{+68}_{-68}$ 		& $1.48^{+0.10}_{-0.12}$ 	& $-0.35^{+0.27}_{-0.27}$ \\
56 & SFG 	& 28 		& $16^{+23}_{-23}$ 		& $89^{+75}_{-75}$ 		& $1.49^{+0.09}_{-0.11}$ 	& $-0.34^{+0.11}_{-0.11}$ \\
57 & SFG 	& 20 		& $2^{+15}_{-15}$ 		& $30^{+38}_{-38}$ 		& $1.60^{+0.08}_{-0.08}$ 	& $-0.30^{+0.26}_{-0.26}$ \\
58 & SFG 	& 32 		& $-24^{+20}_{-20}$ 	& $58^{+68}_{-68}$ 		& $1.44^{+0.04}_{-0.04}$ 	& $-0.23^{+0.16}_{-0.16}$ \\
59 & TG 	& 52 		& $23^{+12}_{-12}$ 		& $23^{+25}_{-26}$ 		& $1.70^{+0.04}_{-0.05}$ 	& $-0.23^{+0.12}_{-0.12}$ \\
60 & SFG 	& 27 		& $-4^{+33}_{-33}$ 		& $71^{+52}_{-53}$ 		& $1.51^{+0.03}_{-0.04}$ 	& $-0.18^{+0.17}_{-0.17}$ \\
61 & TG 	& 30 		& $43^{+12}_{-11}$ 		& $50^{+23}_{-24}$ 		& $1.62^{+0.03}_{-0.04}$ 	& $0.06^{+0.24}_{-0.24}$ \\
62 & SFG 	& 33 		& $21^{+14}_{-14}$ 		& $100^{+29}_{-31}$ 	& $1.60^{+0.03}_{-0.03}$ 	& $0.26^{+0.23}_{-0.23}$ \\
63 & TG 	& 20 		& $47^{+17}_{-17}$ 		& $56^{+56}_{-57}$ 		& $1.52^{+0.02}_{-0.02}$ 	& $0.58^{+0.22}_{-0.22}$ \\
64 & QG 	& 31 		& $15^{+11}_{-11}$ 		& $1^{+22}_{-24}$ 		& $1.84^{+0.10}_{-0.11}$ 	& $0.67^{+0.32}_{-0.32}$ \\
65 & PSB 	& 58 		& $1^{+14}_{-14}$ 		& $1^{+37}_{-38}$ 		& $1.69^{+0.08}_{-0.10}$ 	& $1.03^{+0.17}_{-0.17}$ \\
66 & QG 	& 70 		& $10^{+11}_{-11}$ 		& $2^{+28}_{-29}$ 		& $1.77^{+0.08}_{-0.10}$ 	& $1.04^{+0.22}_{-0.22}$ \\
67 & SFG 	& 25 		& $12^{+11}_{-11}$ 		& $93^{+22}_{-23}$ 		& $1.64^{+0.04}_{-0.05}$ 	& $1.04^{+0.26}_{-0.26}$ \\
68 & QG 	& 82 		& $-4^{+11}_{-11}$ 		& $-11^{+24}_{-25}$ 	& $1.84^{+0.07}_{-0.09}$ 	& $1.13^{+0.48}_{-0.48}$ \\
69 & PSB 	& 59 		& $-35^{+10}_{-10}$ 	& $-27^{+16}_{-18}$		& $1.73^{+0.08}_{-0.08}$ 	& $1.29^{+0.16}_{-0.16}$ \\
70 & QG 	& 67 		& $-4^{+10}_{-10}$ 		& $5^{+13}_{-16}$ 		& $1.75^{+0.11}_{-0.14}$ 	& $1.39^{+0.20}_{-0.20}$ \\
71 & QG 	& 110 		& $13^{+10}_{-10}$ 		& $2^{+14}_{-17}$ 		& $1.78^{+0.02}_{-0.03}$ 	& $1.45^{+0.21}_{-0.21}$
\label{T1}
\end{longtable}
\clearpage

\begin{longtable}{ccrrrrr}
\caption{Parameters Derived From Composite SEDs at $2.5<z<4.0$.}\\
Cid & Class & $N_{gal}$ & $EW_{[OIII]+H\beta}$ (\AA) & \Dfour & $\beta$ \\ \hline \hline \endfirsthead
Cid & Class & $N_{gal}$ & $EW_{[OIII]+H\beta}$ (\AA) & \Dfour & $\beta$ \\ \hline \hline \endhead
0 & ELG 	& 19 		& $2578^{+78}_{-89}$ 	& $0.91^{+0.00}_{-0.00}$ & $-2.05^{+0.07}_{-0.07}$ \\
1 & ELG 	& 64 		& $755^{+5}_{-14}$ 		& $0.94^{+0.08}_{-0.10}$ & $-2.00^{+0.04}_{-0.04}$ \\
2 & SFG 	& 52 		& $336^{+10}_{-11}$ 	& $1.16^{+0.04}_{-0.04}$ & $-1.91^{+0.03}_{-0.03}$ \\
3 & ELG 	& 22 		& $599^{+5}_{-5}$ 		& $0.95^{+0.09}_{-0.11}$ & $-1.87^{+0.13}_{-0.13}$ \\
4 & SFG 	& 89 		& $355^{+7}_{-8}$ 		& $1.17^{+0.03}_{-0.03}$ & $-1.80^{+0.03}_{-0.03}$ \\
5 & SFG 	& 88 		& $224^{+10}_{-16}$ 	& $1.17^{+0.00}_{-0.00}$ & $-1.69^{+0.03}_{-0.03}$ \\
6 & SFG 	& 48 		& $315^{+7}_{-8}$ 		& $1.17^{+0.05}_{-0.06}$ & $-1.66^{+0.05}_{-0.05}$ \\
7 & SFG 	& 37 		& $148^{+9}_{-15}$ 		& $1.23^{+0.01}_{-0.01}$ & $-1.56^{+0.05}_{-0.05}$ \\
8 & SFG 	& 167		& $180^{+9}_{-10}$ 		& $1.19^{+0.02}_{-0.02}$ & $-1.52^{+0.03}_{-0.03}$ \\
9 & SFG 	& 27 		& $100^{+10}_{-10}$ 	& $1.18^{+0.04}_{-0.04}$ & $-1.49^{+0.06}_{-0.06}$ \\
10 & SFG 	& 135 		& $149^{+11}_{-17}$ 	& $1.24^{+0.00}_{-0.00}$ & $-1.35^{+0.04}_{-0.04}$ \\
11 & SFG 	& 76 		& $76^{+10}_{-14}$ 		& $1.23^{+0.00}_{-0.00}$ & $-1.25^{+0.04}_{-0.04}$ \\
12 & SFG 	& 20 		& $1^{+10}_{-11}$ 		& $1.24^{+0.06}_{-0.06}$ & $-1.20^{+0.08}_{-0.08}$ \\
13 & SFG 	& 39 		& $89^{+9}_{-9}$ 		& $1.24^{+0.01}_{-0.01}$ & $-1.11^{+0.05}_{-0.05}$ \\
14 & SFG 	& 39 		& $77^{+10}_{-11}$ 		& $1.28^{+0.00}_{-0.00}$ & $-0.96^{+0.07}_{-0.07}$ \\
15 & SFG 	& 22 		& $57^{+8}_{-10}$ 		& $1.29^{+0.03}_{-0.03}$ & $-0.93^{+0.13}_{-0.13}$ \\
16 & SFG 	& 19 		& $18^{+11}_{-13}$ 		& $1.53^{+0.10}_{-0.11}$ & $-0.70^{+0.42}_{-0.42}$ \\
17 & SFG 	& 30 		& $27^{+12}_{-16}$ 		& $1.36^{+0.02}_{-0.02}$ & $-0.51^{+0.13}_{-0.13}$ \\
18 & PSB 	& 16 		& $38^{+10}_{-12}$ 		& $1.60^{+0.18}_{-0.24}$ & $-0.14^{+0.24}_{-0.24}$ \\
19 & PSB 	& 28 		& $-21^{+10}_{-11}$ 	& $1.71^{+0.09}_{-0.09}$ & $0.49^{+0.28}_{-0.28}$ \\
\label{T2}
\end{longtable}
\clearpage

\begin{longtable}{ccrrrrrrr}
\caption{Analog Galaxy Parameters for Composite SEDs at $1.0<z<3.0$.}\\
Cid & Class & $\log_{10}(M/M_\odot)$ & $sSFR$ (yr$^{-1}$) & \Av\ ($mag$) & \VJc & \UVc & $r_e$ (kpc) & $n$ \\ \hline \hline \endfirsthead
Cid & Class & $\log_{10}(M/M_\odot)$ & $sSFR$ (yr$^{-1}$) & \Av\ ($mag$) & \VJc & \UVc & $r_e$ (kpc) & $n$ \\ \hline \hline \endhead
0 & ELG 	& $8.36^{+0.25}_{-0.23}$	& $-7.00^{+0.03}_{-0.08}$	& $0.40^{+0.10}_{-0.10}$ & $-0.78^{+0.13}_{-0.08}$ & $0.26^{+0.10}_{-0.05}$ & $0.8\pm0.3$ & $2.8\pm1.9$ \\
1 & SFG 	& $9.09^{+0.35}_{-0.45}$ 	& $-8.35^{+1.11}_{-0.37}$ 	& $0.30^{+0.30}_{-0.30}$ & $0.07^{+0.18}_{-0.25}$ & $0.30^{+0.08}_{-0.08}$ & $1.7\pm0.6$ & $1.3\pm0.6$ \\
2 & ELG 	& $8.79^{+0.40}_{-0.27}$ 	& $-7.95^{+0.87}_{-0.93}$ 	& $0.65^{+0.25}_{-0.55}$ & $0.08^{+0.26}_{-0.40}$ & $0.46^{+0.08}_{-0.08}$ & $1.5\pm0.6$ & $1.4\pm0.9$ \\
3 & SFG 	& $9.22^{+0.31}_{-0.38}$ 	& $-8.72^{+0.24}_{-0.27}$ 	& $0.20^{+0.20}_{-0.12}$ & $0.25^{+0.10}_{-0.16}$ & $0.39^{+0.07}_{-0.05}$ & $1.9\pm0.6$ & $1.2\pm0.5$ \\
4 & SFG 	& $9.27^{+0.31}_{-0.32}$ 	& $-8.68^{+0.38}_{-0.32}$ 	& $0.30^{+0.20}_{-0.20}$ & $0.20^{+0.15}_{-0.18}$ & $0.43^{+0.07}_{-0.07}$ & $2.0\pm0.7$ & $1.1\pm0.5$ \\
5 & SFG 	& $9.35^{+0.27}_{-0.35}$ 	& $-8.83^{+0.34}_{-0.21}$ 	& $0.30^{+0.20}_{-0.20}$ & $0.26^{+0.12}_{-0.16}$ & $0.52^{+0.06}_{-0.07}$ & $2.2\pm0.7$ & $1.2\pm0.5$ \\
6 & SFG 	& $9.54^{+0.36}_{-0.24}$ 	& $-9.02^{+0.58}_{-0.24}$ 	& $0.55^{+0.41}_{-0.25}$ & $0.72^{+0.15}_{-0.10}$ & $0.71^{+0.07}_{-0.08}$ & $2.8\pm1.2$ & $1.0\pm0.5$ \\
7 & SFG 	& $9.38^{+0.56}_{-0.24}$ 	& $-8.80^{+0.64}_{-0.40}$ 	& $0.70^{+0.30}_{-0.40}$ & $0.72^{+0.15}_{-0.10}$ & $0.62^{+0.05}_{-0.11}$ & $2.4\pm0.7$ & $1.2\pm0.6$ \\
8 & SFG 	& $9.33^{+0.33}_{-0.40}$ 	& $-8.79^{+0.31}_{-0.41}$ 	& $0.60^{+0.10}_{-0.40}$ & $0.52^{+0.16}_{-0.11}$ & $0.57^{+0.07}_{-0.09}$ & $2.1\pm0.5$ & $0.9\pm0.4$ \\
9 & SFG 	& $9.16^{+0.27}_{-0.23}$ 	& $-8.93^{+0.44}_{-0.29}$ 	& $0.40^{+0.32}_{-0.30}$ & $0.24^{+0.15}_{-0.12}$ & $0.59^{+0.07}_{-0.06}$ & $1.5\pm0.6$ & $1.9\pm0.9$ \\
10 & SFG 	& $9.48^{+0.27}_{-0.31}$ 	& $-8.90^{+0.26}_{-0.30}$ 	& $0.40^{+0.20}_{-0.20}$ & $0.34^{+0.13}_{-0.15}$ & $0.57^{+0.07}_{-0.07}$ & $2.3\pm0.9$ & $1.2\pm0.5$ \\
11 & SFG 	& $9.54^{+0.27}_{-0.34}$ 	& $-8.98^{+0.32}_{-0.22}$ 	& $0.50^{+0.20}_{-0.30}$ & $0.42^{+0.11}_{-0.15}$ & $0.64^{+0.07}_{-0.08}$ & $2.4\pm0.9$ & $1.3\pm0.5$ \\
12 & SFG 	& $9.58^{+0.33}_{-0.39}$ 	& $-9.00^{+0.28}_{-0.23}$ 	& $0.50^{+0.30}_{-0.20}$ & $0.48^{+0.12}_{-0.15}$ & $0.69^{+0.06}_{-0.07}$ & $2.5\pm0.9$ & $1.2\pm0.5$ \\
13 & SFG 	& $9.66^{+0.28}_{-0.26}$ 	& $-8.99^{+0.35}_{-0.22}$ 	& $0.80^{+0.20}_{-0.30}$ & $0.70^{+0.11}_{-0.09}$ & $0.77^{+0.07}_{-0.07}$ & $2.8\pm1.1$ & $1.0\pm0.4$ \\
14 & SFG 	& $9.70^{+0.27}_{-0.32}$ 	& $-9.21^{+0.31}_{-0.43}$ 	& $0.60^{+0.40}_{-0.40}$ & $0.73^{+0.11}_{-0.11}$ & $0.97^{+0.05}_{-0.08}$ & $2.8\pm1.4$ & $1.5\pm0.6$ \\
15 & SFG 	& $9.87^{+0.32}_{-0.28}$ 	& $-8.99^{+0.27}_{-0.21}$ 	& $0.90^{+0.24}_{-0.30}$ & $0.82^{+0.08}_{-0.07}$ & $0.89^{+0.05}_{-0.08}$ & $2.8\pm0.8$ & $1.0\pm0.5$ \\
16 & SFG 	& $10.02^{+0.38}_{-0.42}$ 	& $-8.91^{+0.34}_{-0.28}$ 	& $1.00^{+0.30}_{-0.20}$ & $1.02^{+0.07}_{-0.13}$ & $0.86^{+0.05}_{-0.07}$ & $3.3\pm1.2$ & $0.9\pm0.5$ \\
17 & SFG 	& $9.64^{+0.34}_{-0.30}$ 	& $-9.00^{+0.23}_{-0.40}$ 	& $0.70^{+0.20}_{-0.30}$ & $0.53^{+0.11}_{-0.14}$ & $0.75^{+0.06}_{-0.08}$ & $2.4\pm1.0$ & $1.4\pm0.6$ \\
18 & SFG 	& $9.88^{+0.26}_{-0.40}$ 	& $-8.99^{+0.31}_{-0.21}$ 	& $0.80^{+0.30}_{-0.20}$ & $0.83^{+0.11}_{-0.09}$ & $0.82^{+0.06}_{-0.06}$ & $3.4\pm1.2$ & $1.0\pm0.4$ \\
19 & SFG 	& $10.22^{+0.26}_{-0.58}$ 	& $-9.12^{+0.50}_{-0.23}$ 	& $1.10^{+0.42}_{-0.20}$ & $1.17^{+0.12}_{-0.05}$ & $0.99^{+0.07}_{-0.08}$ & $4.0\pm0.9$ & $0.7\pm0.2$ \\
20 & SFG 	& $9.79^{+0.35}_{-0.45}$ 	& $-9.41^{+0.20}_{-0.33}$ 	& $0.60^{+0.20}_{-0.40}$ & $0.78^{+0.05}_{-0.05}$ & $1.06^{+0.06}_{-0.04}$ & $2.2\pm0.5$ & $1.6\pm0.7$ \\
21 & SFG 	& $9.79^{+0.28}_{-0.35}$ 	& $-8.99^{+0.27}_{-0.22}$ 	& $0.90^{+0.10}_{-0.30}$ & $0.68^{+0.10}_{-0.10}$ & $0.83^{+0.06}_{-0.07}$ & $2.9\pm1.0$ & $1.1\pm0.4$ \\
22 & SFG 	& $10.00^{+0.35}_{-0.34}$ 	& $-8.99^{+0.21}_{-0.25}$ 	& $1.15^{+0.15}_{-0.25}$ & $0.99^{+0.07}_{-0.06}$ & $0.96^{+0.05}_{-0.08}$ & $3.4\pm0.9$ & $1.1\pm0.4$ \\
23 & SFG 	& $9.70^{+0.27}_{-0.37}$ 	& $-9.17^{+0.27}_{-0.37}$ 	& $0.50^{+0.20}_{-0.20}$ & $0.38^{+0.11}_{-0.12}$ & $0.68^{+0.05}_{-0.07}$ & $2.2\pm0.8$ & $1.3\pm0.5$ \\
24 & SFG 	& $9.96^{+0.30}_{-0.37}$ 	& $-9.00^{+0.19}_{-0.23}$ 	& $1.00^{+0.20}_{-0.30}$ & $0.84^{+0.09}_{-0.09}$ & $0.93^{+0.06}_{-0.06}$ & $3.3\pm1.2$ & $1.2\pm0.6$ \\
25 & SFG 	& $10.19^{+0.43}_{-0.62}$ 	& $-9.93^{+0.37}_{-0.18}$ 	& $0.60^{+0.58}_{-0.30}$ & $0.99^{+0.08}_{-0.07}$ & $1.33^{+0.06}_{-0.05}$ & $1.8\pm0.9$ & $2.4\pm0.7$ \\
26 & SFG 	& $9.81^{+0.28}_{-0.42}$ 	& $-9.17^{+0.36}_{-0.43}$ 	& $0.70^{+0.20}_{-0.20}$ & $0.57^{+0.10}_{-0.14}$ & $0.75^{+0.07}_{-0.06}$ & $2.4\pm0.7$ & $1.3\pm0.6$ \\
27 & SFG 	& $10.27^{+0.22}_{-0.42}$ 	& $-9.20^{+0.27}_{-0.44}$ 	& $1.40^{+0.20}_{-0.40}$ & $1.08^{+0.08}_{-0.06}$ & $1.14^{+0.06}_{-0.06}$ & $3.4\pm0.9$ & $0.9\pm0.4$ \\
28 & SFG 	& $10.71^{+0.31}_{-0.33}$ 	& $-10.36^{+0.31}_{-1.56}$ 	& $1.75^{+0.85}_{-0.25}$ & $1.88^{+0.12}_{-0.06}$ & $2.01^{+0.14}_{-0.09}$ & $3.4\pm0.8$ & $1.4\pm0.6$ \\
29 & SFG 	& $10.39^{+0.33}_{-0.47}$ 	& $-9.28^{+0.29}_{-0.18}$ 	& $1.30^{+0.40}_{-0.20}$ & $1.36^{+0.04}_{-0.04}$ & $1.20^{+0.03}_{-0.10}$ & $3.6\pm1.4$ & $0.8\pm0.2$ \\
30 & SFG 	& $10.03^{+0.31}_{-0.42}$ 	& $-9.04^{+0.20}_{-0.40}$ 	& $1.10^{+0.20}_{-0.30}$ & $0.93^{+0.07}_{-0.10}$ & $1.02^{+0.05}_{-0.06}$ & $3.1\pm1.3$ & $1.1\pm0.6$ \\
31 & SFG 	& $10.01^{+0.36}_{-0.30}$ 	& $-9.36^{+0.30}_{-0.29}$ 	& $1.20^{+0.20}_{-0.40}$ & $0.92^{+0.05}_{-0.09}$ & $1.15^{+0.02}_{-0.05}$ & $2.2\pm0.8$ & $1.2\pm0.4$ \\
32 & SFG 	& $9.89^{+0.33}_{-0.37}$ 	& $-9.08^{+0.18}_{-0.36}$ 	& $0.90^{+0.20}_{-0.20}$ & $0.65^{+0.10}_{-0.13}$ & $0.85^{+0.06}_{-0.06}$ & $2.6\pm0.9$ & $1.1\pm0.5$ \\
33 & SFG 	& $10.26^{+0.29}_{-0.42}$ 	& $-9.19^{+0.40}_{-0.45}$ 	& $1.70^{+0.20}_{-0.33}$ & $1.36^{+0.10}_{-0.04}$ & $1.22^{+0.09}_{-0.04}$ & $3.5\pm0.8$ & $0.9\pm0.3$ \\
34 & SFG 	& $10.11^{+0.27}_{-0.40}$ 	& $-9.09^{+0.28}_{-0.29}$ 	& $1.30^{+0.18}_{-0.30}$ & $1.04^{+0.08}_{-0.05}$ & $1.02^{+0.09}_{-0.09}$ & $3.6\pm0.8$ & $0.9\pm0.5$ \\
35 & SFG 	& $10.06^{+0.50}_{-0.33}$ 	& $-9.09^{+0.31}_{-0.32}$ 	& $1.55^{+0.15}_{-0.45}$ & $1.25^{+0.08}_{-0.09}$ & $1.13^{+0.08}_{-0.07}$ & $3.6\pm1.0$ & $0.9\pm0.3$ \\
36 & SFG 	& $9.73^{+0.42}_{-0.40}$ 	& $-9.36^{+0.36}_{-0.79}$ 	& $0.60^{+0.20}_{-0.10}$ & $0.48^{+0.11}_{-0.11}$ & $0.77^{+0.07}_{-0.08}$ & $2.3\pm0.8$ & $1.4\pm0.5$ \\
37 & SFG 	& $9.91^{+0.38}_{-0.48}$ 	& $-9.36^{+0.33}_{-0.41}$ 	& $0.80^{+0.20}_{-0.22}$ & $0.60^{+0.08}_{-0.08}$ & $0.89^{+0.07}_{-0.06}$ & $2.4\pm0.9$ & $1.6\pm0.7$ \\
38 & SFG 	& $10.45^{+0.33}_{-0.45}$ 	& $-9.18^{+0.37}_{-0.30}$ 	& $1.65^{+0.35}_{-0.35}$ & $1.41^{+0.07}_{-0.07}$ & $1.32^{+0.06}_{-0.06}$ & $3.7\pm1.1$ & $1.1\pm0.5$ \\
39 & SFG 	& $10.29^{+0.32}_{-0.41}$ 	& $-9.12^{+0.34}_{-0.28}$ 	& $1.55^{+0.25}_{-0.45}$ & $1.27^{+0.08}_{-0.07}$ & $1.12^{+0.09}_{-0.05}$ & $3.3\pm0.5$ & $0.8\pm0.1$ \\
40 & TG 	& $10.51^{+0.58}_{-0.40}$ 	& $-10.09^{+0.17}_{-0.63}$ 	& $0.70^{+0.38}_{-0.50}$ & $1.19^{+0.05}_{-0.05}$ & $1.60^{+0.07}_{-0.07}$ & $2.2\pm1.1$ & $4.6\pm2.4$ \\
41 & SFG 	& $10.02^{+0.36}_{-0.40}$ 	& $-9.17^{+0.26}_{-1.05}$ 	& $1.30^{+0.10}_{-0.20}$ & $0.96^{+0.04}_{-0.06}$ & $1.05^{+0.05}_{-0.06}$ & $2.7\pm1.2$ & $1.4\pm0.6$ \\
42 & SFG 	& $10.00^{+0.29}_{-0.28}$ 	& $-9.00^{+0.34}_{-0.77}$ 	& $1.10^{+0.20}_{-0.10}$ & $0.83^{+0.08}_{-0.08}$ & $0.89^{+0.03}_{-0.08}$ & $2.9\pm1.0$ & $0.8\pm0.3$ \\
43 & SFG 	& $9.92^{+0.22}_{-0.33}$ 	& $-9.20^{+0.20}_{-0.52}$ 	& $0.90^{+0.20}_{-0.20}$ & $0.74^{+0.10}_{-0.06}$ & $0.97^{+0.05}_{-0.09}$ & $2.7\pm1.2$ & $1.5\pm0.7$ \\
44 & SFG 	& $10.10^{+0.57}_{-0.44}$ 	& $-9.44^{+0.46}_{-0.44}$ 	& $1.40^{+0.20}_{-0.60}$ & $1.08^{+0.09}_{-0.05}$ & $1.20^{+0.08}_{-0.06}$ & $2.7\pm0.9$ & $0.9\pm0.3$ \\
45 & SFG 	& $9.67^{+0.51}_{-0.47}$ 	& $-10.46^{+0.95}_{-1.80}$ 	& $0.70^{+0.10}_{-0.28}$ & $0.46^{+0.09}_{-0.15}$ & $0.82^{+0.05}_{-0.06}$ & $2.1\pm0.7$ & $2.6\pm0.9$ \\
46 & SFG 	& $10.05^{+0.26}_{-0.44}$ 	& $-9.36^{+0.36}_{-0.63}$ 	& $1.20^{+0.10}_{-0.20}$ & $0.85^{+0.08}_{-0.12}$ & $1.01^{+0.07}_{-0.03}$ & $2.5\pm0.8$ & $1.2\pm0.5$ \\
47 & SFG 	& $10.32^{+0.34}_{-0.35}$ 	& $-9.23^{+0.24}_{-0.53}$ 	& $1.60^{+0.20}_{-0.40}$ & $1.26^{+0.05}_{-0.08}$ & $1.28^{+0.05}_{-0.06}$ & $3.0\pm1.3$ & $1.3\pm0.7$ \\
48 & SFG 	& $10.43^{+0.38}_{-0.40}$ 	& $-9.58^{+0.39}_{-0.71}$ 	& $1.60^{+0.30}_{-0.60}$ & $1.28^{+0.09}_{-0.07}$ & $1.39^{+0.09}_{-0.06}$ & $2.8\pm0.8$ & $1.6\pm0.6$ \\
49 & SFG 	& $10.59^{+0.38}_{-0.19}$ 	& $-9.17^{+0.39}_{-0.44}$ 	& $2.60^{+0.17}_{-0.60}$ & $1.92^{+0.07}_{-0.04}$ & $1.61^{+0.09}_{-0.09}$ & $4.5\pm0.9$ & $0.8\pm0.3$ \\
50 & SFG 	& $9.95^{+0.49}_{-0.48}$ 	& $-9.58^{+0.39}_{-0.88}$ 	& $1.40^{+0.10}_{-0.60}$ & $0.99^{+0.07}_{-0.16}$ & $1.22^{+0.10}_{-0.05}$ & $2.0\pm1.0$ & $2.1\pm0.7$ \\
51 & SFG 	& $10.36^{+0.43}_{-0.58}$	& $-9.58^{+0.21}_{-1.59}$ 	& $1.50^{+0.20}_{-0.30}$ & $1.10^{+0.08}_{-0.06}$ & $1.33^{+0.07}_{-0.05}$ & $2.4\pm0.9$ & $2.1\pm0.8$ \\
52 & TG 	& $10.23^{+0.35}_{-0.57}$ 	& $-10.36^{+0.44}_{-0.32}$ 	& $0.80^{+0.60}_{-0.55}$ & $1.00^{+0.04}_{-0.05}$ & $1.44^{+0.06}_{-0.05}$ & $1.3\pm0.6$ & $3.7\pm2.4$ \\
53 & SFG 	& $10.70^{+0.28}_{-0.32}$ 	& $-9.74^{+0.74}_{-1.12}$ 	& $2.80^{+0.10}_{-0.66}$ & $1.99^{+0.07}_{-0.06}$ & $1.76^{+0.09}_{-0.07}$ & $3.8\pm0.8$ & $1.2\pm0.4$ \\
54 & TG 	& $10.41^{+0.37}_{-0.42}$ 	& $-10.15^{+0.24}_{-0.40}$ 	& $1.00^{+0.50}_{-0.71}$ & $1.12^{+0.06}_{-0.08}$ & $1.52^{+0.05}_{-0.07}$ & $2.3\pm1.0$ & $3.0\pm1.0$ \\
55 & SFG 	& $10.79^{+0.35}_{-0.32}$ 	& $-9.77^{+0.49}_{-0.61}$ 	& $2.90^{+0.30}_{-0.60}$ & $2.18^{+0.14}_{-0.06}$ & $1.88^{+0.13}_{-0.10}$ & $4.1\pm0.9$ & $1.3\pm0.4$ \\
56 & TG 	& $10.57^{+0.26}_{-0.35}$ 	& $-9.84^{+0.43}_{-0.38}$ 	& $1.45^{+0.62}_{-0.62}$ & $1.41^{+0.09}_{-0.07}$ & $1.56^{+0.05}_{-0.09}$ & $3.4\pm1.3$ & $2.0\pm0.5$ \\
57 & SFG 	& $10.64^{+0.25}_{-0.18}$ 	& $-9.84^{+0.39}_{-0.86}$ 	& $1.90^{+0.50}_{-0.70}$ & $1.64^{+0.04}_{-0.09}$ & $1.74^{+0.04}_{-0.09}$ & $3.9\pm1.5$ & $1.7\pm0.7$ \\
58 & SFG 	& $10.37^{+0.43}_{-0.30}$ 	& $-9.45^{+0.45}_{-0.33}$ 	& $2.00^{+0.10}_{-0.70}$ & $1.48^{+0.05}_{-0.04}$ & $1.43^{+0.08}_{-0.05}$ & $3.3\pm1.4$ & $1.1\pm0.4$ \\
59 & TG 	& $10.51^{+0.41}_{-0.37}$ 	& $-10.70^{+0.34}_{-0.38}$ 	& $0.40^{+0.40}_{-0.30}$ & $1.08^{+0.10}_{-0.04}$ & $1.65^{+0.08}_{-0.06}$ & $1.7\pm1.0$ & $4.7\pm1.7$ \\
60 & SFG 	& $10.50^{+0.31}_{-0.23}$ 	& $-9.56^{+0.28}_{-1.74}$ 	& $2.00^{+0.30}_{-0.30}$ & $1.54^{+0.08}_{-0.05}$ & $1.56^{+0.10}_{-0.07}$ & $4.3\pm1.6$ & $1.6\pm0.6$ \\
61 & TG 	& $10.64^{+0.19}_{-0.46}$ 	& $-10.41^{+0.05}_{-0.37}$ 	& $0.85^{+0.52}_{-0.19}$ & $1.27^{+0.04}_{-0.06}$ & $1.68^{+0.05}_{-0.07}$ & $2.1\pm0.8$ & $2.2\pm1.1$ \\
62 & TG 	& $10.75^{+0.27}_{-0.33}$ 	& $-10.15^{+0.43}_{-1.01}$ 	& $2.10^{+0.39}_{-0.80}$ & $1.75^{+0.07}_{-0.06}$ & $1.84^{+0.09}_{-0.11}$ & $3.2\pm1.1$ & $1.3\pm0.6$ \\
63 & TG 	& $10.50^{+0.29}_{-0.57}$ 	& $-10.91^{+0.75}_{-0.65}$ 	& $0.95^{+0.45}_{-0.84}$ & $1.08^{+0.05}_{-0.08}$ & $1.52^{+0.05}_{-0.06}$ & $1.9\pm0.6$ & $3.5\pm1.8$ \\
64 & QG 	& $10.85^{+0.20}_{-0.31}$ 	& $-11.08^{+0.54}_{-0.54}$ 	& $0.80^{+0.62}_{-0.20}$ & $1.48^{+0.06}_{-0.12}$ & $2.01^{+0.10}_{-0.13}$ & $2.3\pm0.7$ & $2.9\pm1.1$ \\
65 & PSB	& $10.52^{+0.34}_{-0.43}$ 	& $-11.23^{+0.53}_{-1.78}$ 	& $0.40^{+0.10}_{-0.20}$ & $0.82^{+0.07}_{-0.11}$ & $1.54^{+0.04}_{-0.04}$ & $0.9\pm0.3$ & $3.5\pm1.1$ \\
66 & QG 	& $10.57^{+0.38}_{-0.31}$ 	& $-11.38^{+0.47}_{-1.60}$ 	& $0.50^{+0.20}_{-0.20}$ & $1.16^{+0.05}_{-0.04}$ & $1.83^{+0.06}_{-0.06}$ & $1.7\pm0.6$ & $4.5\pm1.3$ \\
67 & TG 	& $10.51^{+0.42}_{-0.36}$ 	& $-10.54^{+0.39}_{-3.10}$ 	& $1.10^{+0.82}_{-0.22}$ & $1.40^{+0.10}_{-0.06}$ & $1.67^{+0.15}_{-0.06}$ & $2.8\pm1.2$ & $1.5\pm0.7$ \\
68 & QG 	& $10.68^{+0.27}_{-0.37}$ 	& $-11.69^{+0.31}_{-1.55}$ 	& $0.50^{+0.20}_{-0.30}$ & $1.26^{+0.07}_{-0.06}$ & $1.91^{+0.06}_{-0.05}$ & $1.7\pm0.6$ & $3.5\pm1.1$ \\
69 & PSB 	& $10.51^{+0.32}_{-0.35}$ 	& $-11.38^{+0.62}_{-4.10}$ 	& $0.50^{+0.17}_{-0.40}$ & $0.93^{+0.09}_{-0.06}$ & $1.60^{+0.05}_{-0.04}$ & $1.1\pm0.4$ & $3.6\pm1.1$ \\
70 & QG 	& $10.37^{+0.47}_{-0.43}$ 	& $-11.38^{+0.15}_{-5.47}$ 	& $0.30^{+0.44}_{-0.10}$ & $1.00^{+0.05}_{-0.05}$ & $1.69^{+0.07}_{-0.04}$ & $1.3\pm0.4$ & $3.7\pm0.9$ \\
71 & QG 	& $10.57^{+0.31}_{-0.28}$ 	& $-11.62^{+0.24}_{-2.76}$ 	& $0.40^{+0.30}_{-0.20}$ & $1.09^{+0.05}_{-0.03}$ & $1.77^{+0.05}_{-0.05}$ & $1.4\pm0.5$ & $3.8\pm1.1$ \\
\label{T3}
\end{longtable}
\clearpage

\begin{longtable}{ccrrrrrrr}
\caption{Analog Galaxy Parameters for Composite SEDs at $2.5<z<4.0$.}\\
Cid & Class & $\log_{10}(M/M_\odot)$ & $sSFR$ (yr$^{-1}$) & \Av\ ($mag$) & \VJc & \UVc & $r_e$ (kpc) & $n$ \\ \hline \hline \endfirsthead
Cid & Class & $\log_{10}(M/M_\odot)$ & $sSFR$ (yr$^{-1}$) & \Av\ ($mag$) & \VJc & \UVc & $r_e$ (kpc) & $n$ \\ \hline \hline \endhead
0 & ELG & $8.59^{+0.63}_{-0.33}$ & $-6.87^{+0.15}_{-1.51}$ & $0.70^{+0.11}_{-0.70}$ & $-0.36^{+0.38}_{-0.24}$ & $0.35^{+0.19}_{-0.25}$ & $1.1\pm0.5$ & $1.5\pm1.2$ \\
1 & ELG & $8.84^{+0.49}_{-0.24}$ & $-7.13^{+0.10}_{-1.25}$ & $0.60^{+0.10}_{-0.50}$ & $-0.18^{+0.35}_{-0.34}$ & $0.24^{+0.17}_{-0.09}$ & $1.2\pm0.5$ & $1.3\pm0.8$ \\
2 & SFG & $9.30^{+0.23}_{-0.43}$ & $-8.45^{+1.25}_{-0.33}$ & $0.15^{+0.43}_{-0.15}$ & $-0.07^{+0.30}_{-0.30}$ & $0.23^{+0.08}_{-0.05}$ & $1.5\pm0.4$ & $1.6\pm0.9$ \\
3 & ELG & $8.84^{+0.26}_{-0.27}$ & $-7.06^{+0.09}_{-0.08}$ & $0.90^{+0.10}_{-0.10}$ & $-0.10^{+0.16}_{-0.25}$ & $0.35^{+0.23}_{-0.07}$ & $1.5\pm0.6$ & $1.2\pm0.8$ \\
4 & SFG & $9.43^{+0.30}_{-0.54}$ & $-8.30^{+1.10}_{-0.60}$ & $0.30^{+0.50}_{-0.30}$ & $0.06^{+0.30}_{-0.17}$ & $0.30^{+0.11}_{-0.10}$ & $1.6\pm0.6$ & $1.5\pm0.7$ \\
5 & SFG & $9.64^{+0.19}_{-0.29}$ & $-8.99^{+0.46}_{-0.18}$ & $0.20^{+0.20}_{-0.20}$ & $0.11^{+0.21}_{-0.23}$ & $0.35^{+0.16}_{-0.11}$ & $1.6\pm0.7$ & $1.1\pm0.5$ \\
6 & SFG & $9.12^{+0.72}_{-0.35}$ & $-7.24^{+0.09}_{-1.71}$ & $0.80^{+0.20}_{-0.70}$ & $0.06^{+0.32}_{-0.31}$ & $0.36^{+0.12}_{-0.09}$ & $1.7\pm0.5$ & $1.0\pm0.6$ \\
7 & SFG & $9.76^{+0.19}_{-0.24}$ & $-9.00^{+0.36}_{-0.36}$ & $0.30^{+0.32}_{-0.30}$ & $0.21^{+0.30}_{-0.13}$ & $0.55^{+0.09}_{-0.16}$ & $1.7\pm0.6$ & $1.9\pm1.0$ \\
8 & SFG & $9.72^{+0.24}_{-0.31}$ & $-8.99^{+0.64}_{-0.37}$ & $0.30^{+0.30}_{-0.20}$ & $0.22^{+0.25}_{-0.23}$ & $0.41^{+0.14}_{-0.10}$ & $1.7\pm0.5$ & $1.2\pm0.6$ \\
9 & SFG & $9.97^{+0.25}_{-0.31}$ & $-8.82^{+0.29}_{-0.21}$ & $0.60^{+0.28}_{-0.10}$ & $0.58^{+0.28}_{-0.20}$ & $0.62^{+0.11}_{-0.08}$ & $2.1\pm0.8$ & $1.1\pm0.8$ \\
10 & SFG & $9.83^{+0.26}_{-0.16}$ & $-9.00^{+0.21}_{-0.44}$ & $0.40^{+0.20}_{-0.30}$ & $0.30^{+0.20}_{-0.19}$ & $0.51^{+0.13}_{-0.12}$ & $1.8\pm0.7$ & $1.3\pm0.7$ \\
11 & SFG & $10.02^{+0.25}_{-0.27}$ & $-9.00^{+0.19}_{-0.44}$ & $0.50^{+0.30}_{-0.20}$ & $0.42^{+0.21}_{-0.16}$ & $0.62^{+0.10}_{-0.14}$ & $2.0\pm0.6$ & $1.4\pm0.8$ \\
12 & SFG & $10.23^{+0.27}_{-0.30}$ & $-8.87^{+0.22}_{-0.20}$ & $1.00^{+0.10}_{-0.10}$ & $0.76^{+0.22}_{-0.04}$ & $0.83^{+0.09}_{-0.11}$ & $2.2\pm1.1$ & $1.1\pm0.3$ \\
13 & SFG & $10.08^{+0.23}_{-0.26}$ & $-9.07^{+0.41}_{-0.52}$ & $0.70^{+0.20}_{-0.30}$ & $0.54^{+0.21}_{-0.28}$ & $0.65^{+0.16}_{-0.17}$ & $2.1\pm0.6$ & $1.5\pm0.5$ \\
14 & SFG & $10.05^{+0.22}_{-0.20}$ & $-9.44^{+0.45}_{-0.55}$ & $0.70^{+0.29}_{-0.30}$ & $0.49^{+0.20}_{-0.15}$ & $0.70^{+0.07}_{-0.15}$ & $2.8\pm1.1$ & $1.1\pm0.4$ \\
15 & SFG & $9.99^{+0.30}_{-0.26}$ & $-9.61^{+0.58}_{-1.56}$ & $0.80^{+0.10}_{-0.20}$ & $0.55^{+0.12}_{-0.19}$ & $0.84^{+0.04}_{-0.16}$ & $1.8\pm0.8$ & $1.1\pm0.6$ \\
16 & SFG & $10.96^{+0.20}_{-0.50}$ & $-9.74^{+1.30}_{-0.40}$ & $2.50^{+0.22}_{-0.41}$ & $1.91^{+0.27}_{-0.11}$ & $1.64^{+0.31}_{-0.30}$ & $2.8\pm1.2$ & $0.9\pm0.4$ \\
17 & SFG & $10.53^{+0.23}_{-0.41}$ & $-9.07^{+0.55}_{-0.89}$ & $1.50^{+0.34}_{-0.20}$ & $1.30^{+0.33}_{-0.15}$ & $1.15^{+0.16}_{-0.15}$ & $2.8\pm0.7$ & $1.2\pm0.9$ \\
18 & PSB & $10.50^{+0.39}_{-0.33}$ & $-10.36^{+0.31}_{-0.69}$ & $0.40^{+0.46}_{-0.30}$ & $0.78^{+0.07}_{-0.35}$ & $1.40^{+0.09}_{-0.05}$ & $1.1\pm0.7$ & $7.8\pm0.2$ \\
19 & PSB & $10.66^{+0.34}_{-0.20}$ & $-11.66^{+0.96}_{-2.64}$ & $0.50^{+0.30}_{-0.10}$ & $0.89^{+0.21}_{-0.18}$ & $1.59^{+0.22}_{-0.09}$ & $0.9\pm0.4$ & $3.1\pm1.2$ \\
\label{T4}
\end{longtable}
\clearpage



	\begin{figure}[tp]
	\includegraphics[width=0.61\textwidth,trim=0in 2in 0in 0in, clip=true]{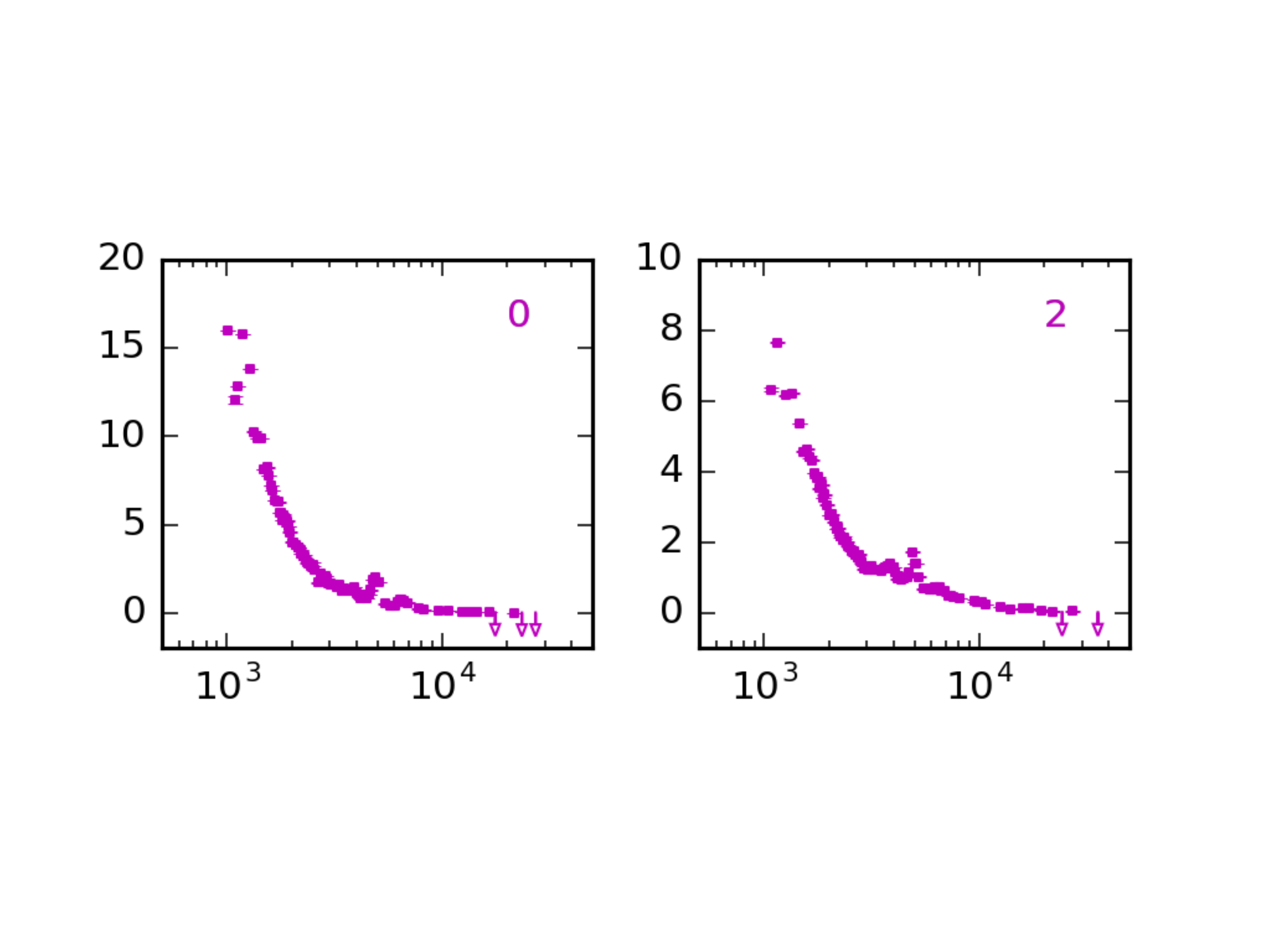}
	\caption{The set of ELG composite SEDs from $1<z<3$ as scaled $F_\lambda$ against wavelength.}
	\label{fig:ELG_lo}
	\end{figure}



	\begin{figure}[tp]
	\includegraphics[width=0.92\textwidth,trim=0in 3in 0in 0in, clip=true]{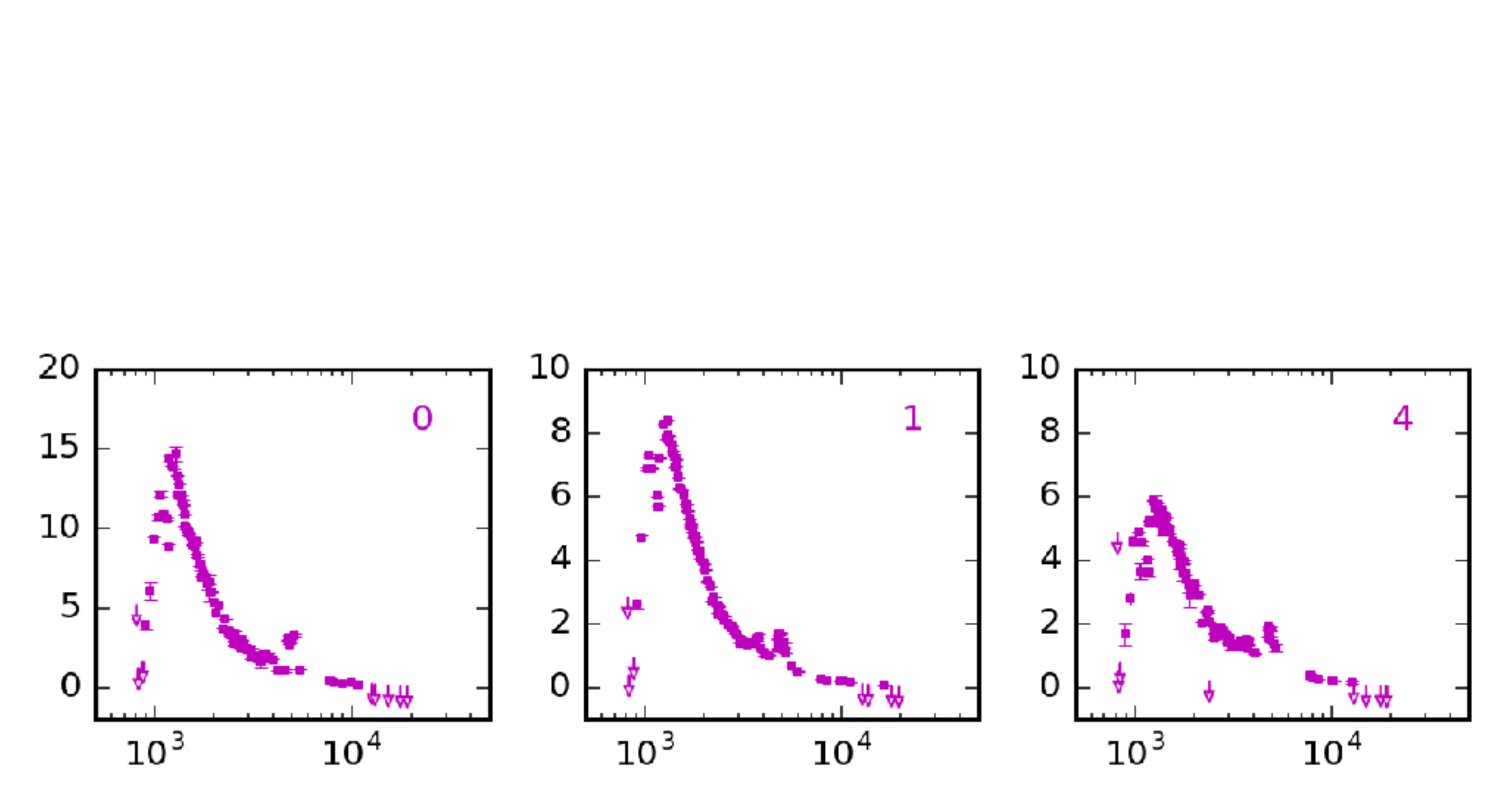}
	\caption{The set of ELG composite SEDs from $2.5<z<4$ as scaled $F_\lambda$ against wavelength.}
	\label{fig:ELG_hi}
	\end{figure}



	\begin{figure}[tp]
	\centerline{\includegraphics[width=0.85\textwidth,trim=0in 0in 0in 0in, clip=true]{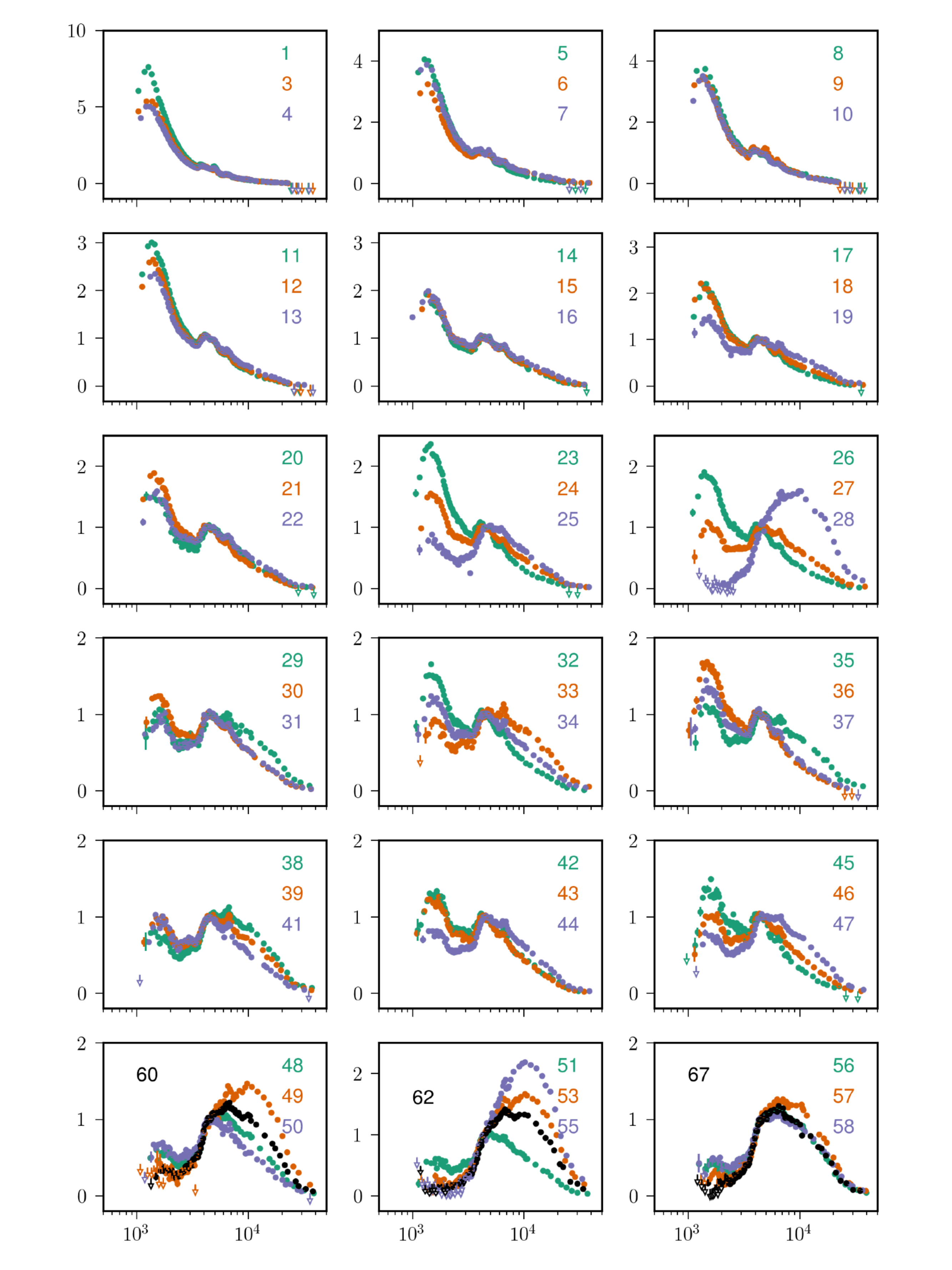}}
	\caption{The set of SFG composite SEDs from $1<z<3$ as scaled $F_\lambda$ against wavelength.  Each panel has 3 composite SEDs-- we have changed the blue color used throughout much of the paper for clarity and grouped by UV flux. Note that the y-axis range changes between panels, although the abscissae are identical. There are many composite SED pairs which are similar, and we do not claim that these are all separate populations, although many have differences, as shown throughout this work.}
	\label{fig:SFG_lo}
	\end{figure}



	\begin{figure}[tp]
	\centerline{\includegraphics[width=0.85\textwidth,trim=0in 1in 0in 0in, clip=true]{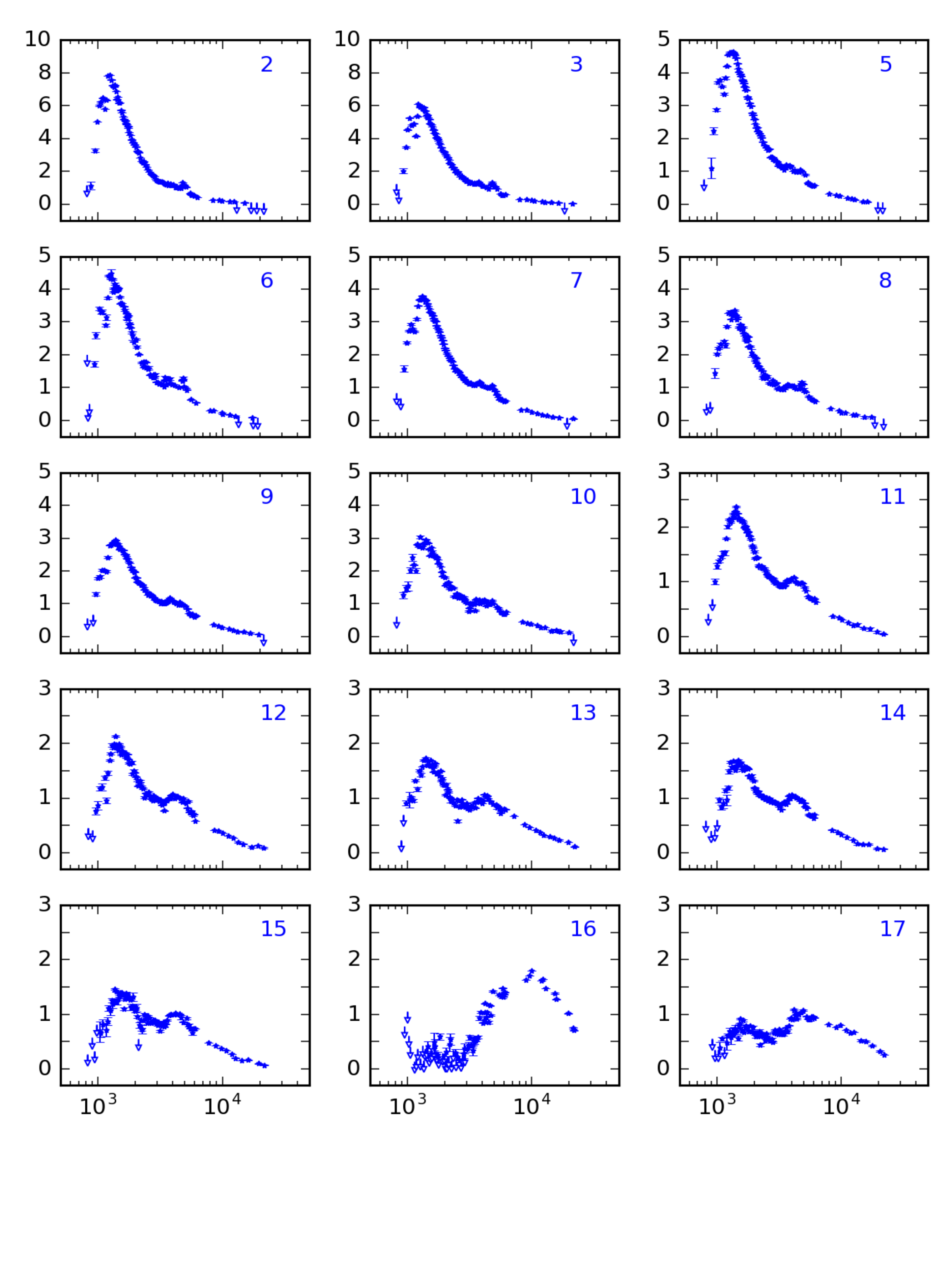}}
	\caption{The set of SFG composite SEDs from $2.5<z<4$ as scaled $F_\lambda$ against wavelength. Note that the y-axis range changes between panels, although the abscissae are identical.}
	\label{fig:SFG_hi}
	\end{figure}



	\begin{figure}[tp]
	\centerline{\includegraphics[width=0.92\textwidth,trim=0in 1in 0in 0in, clip=true]{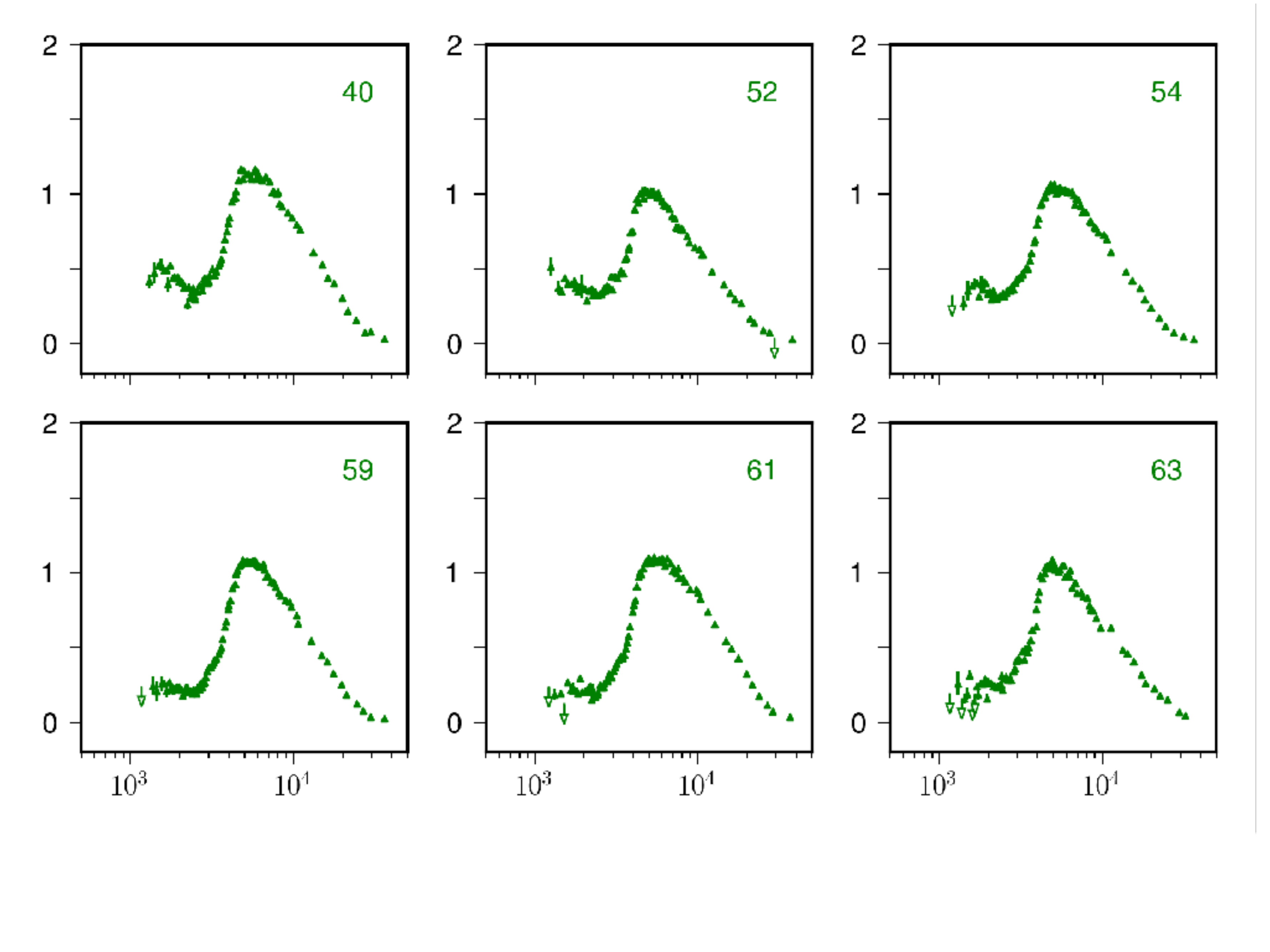}}
	\caption{The set of TG composite SEDs from $1<z<3$ as scaled $F_\lambda$ against wavelength.}
	\label{fig:TG_lo}
	\end{figure}


\clearpage

	\begin{figure}[tp]
	\includegraphics[width=0.61\textwidth,trim=0in 3in 0in 0in, clip=true]{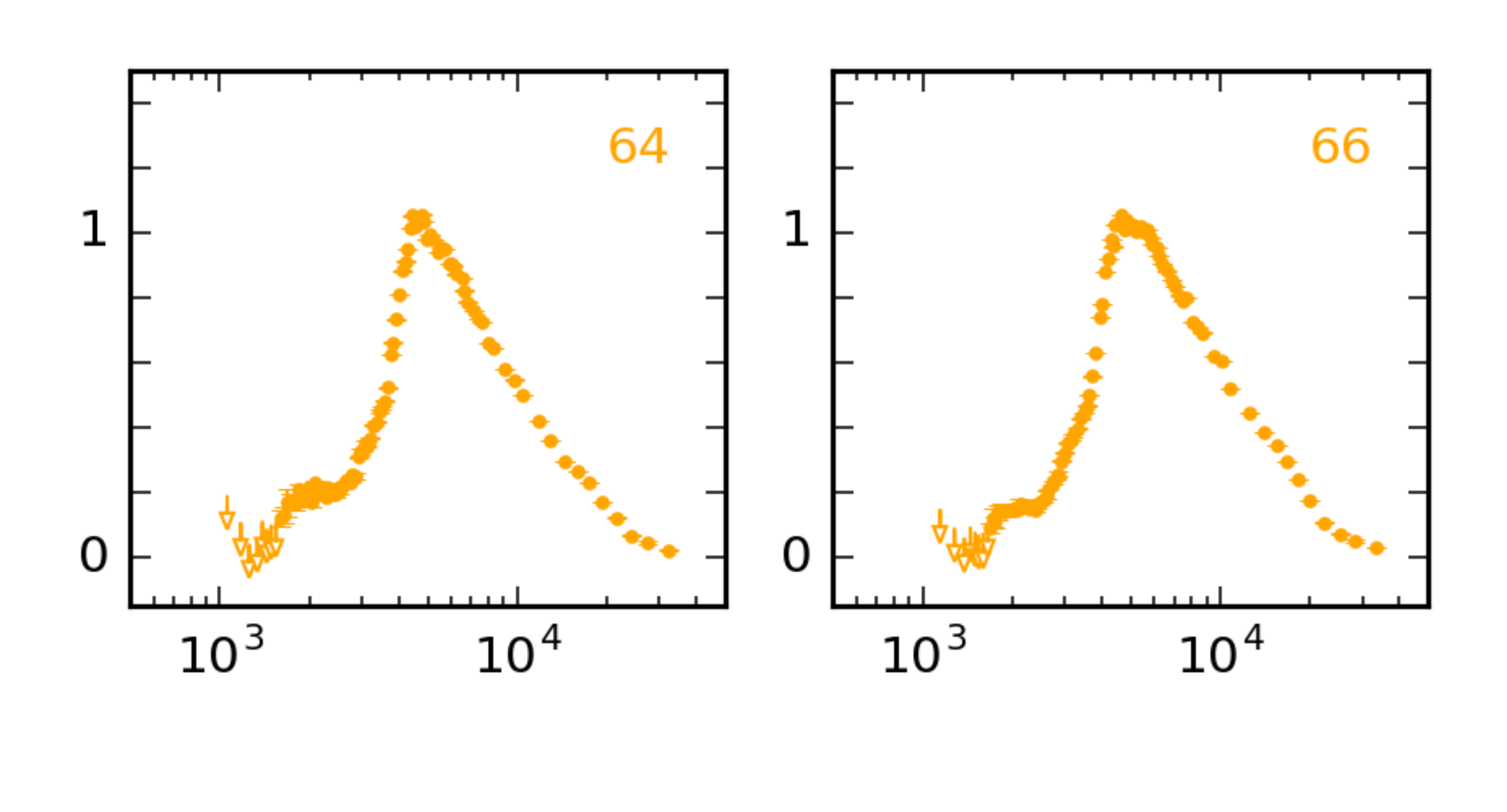}
	\caption{The set of PSB composite SEDs from $1<z<3$ as scaled $F_\lambda$ against wavelength.}
	\label{fig:PSB_lo}
	\end{figure}



	\begin{figure}[tp]
	\includegraphics[width=0.61\textwidth,trim=0in 3in 0in 0in, clip=true]{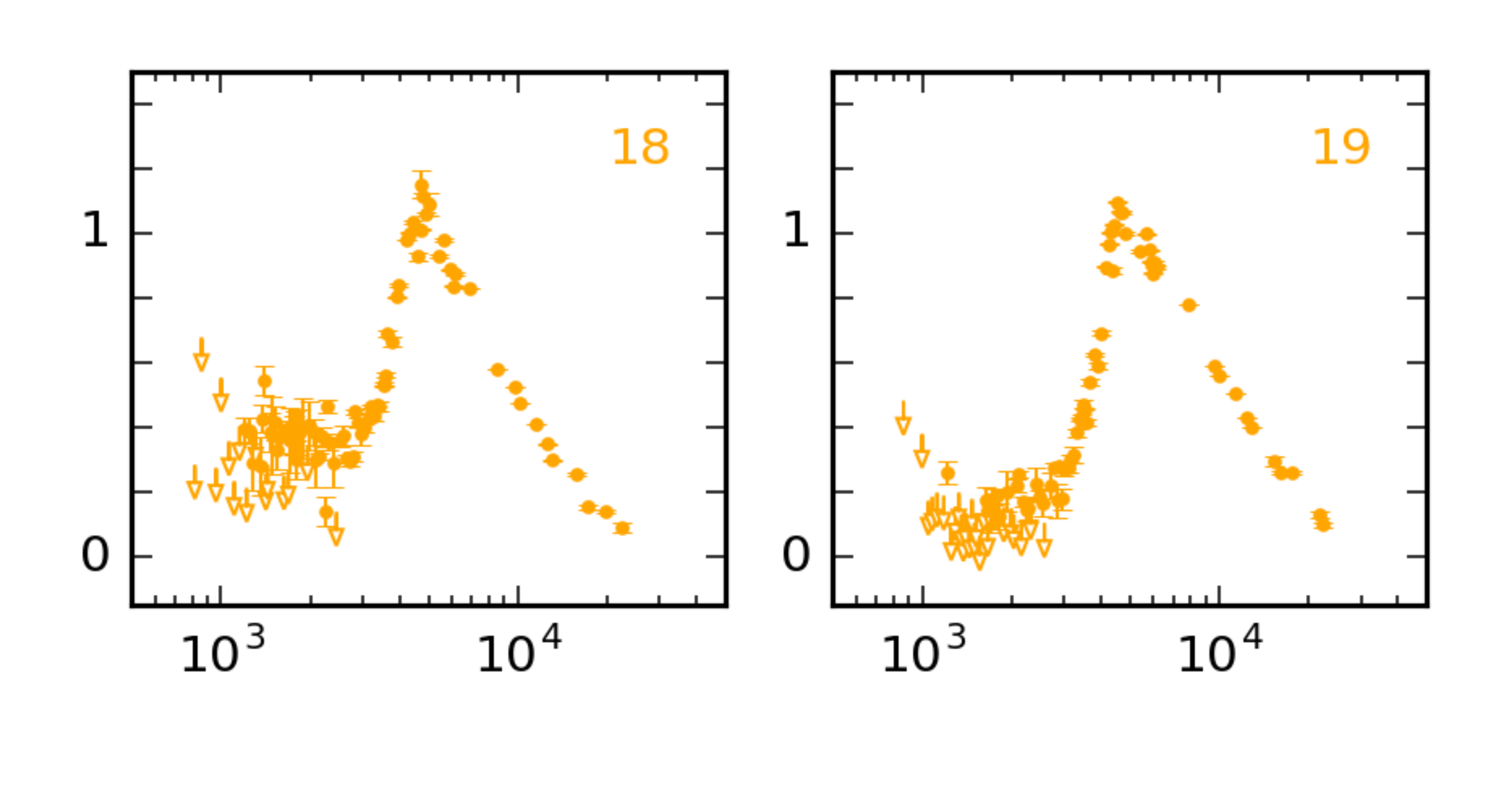}
	\caption{The set of PSB composite SEDs from $2.5<z<4$ as scaled $F_\lambda$ against wavelength.}
	\label{fig:PSB_hi}
	\end{figure}



	\begin{figure}[tp]
	\centerline{\includegraphics[width=0.92\textwidth,trim=0in 1in 0in 0in, clip=true]{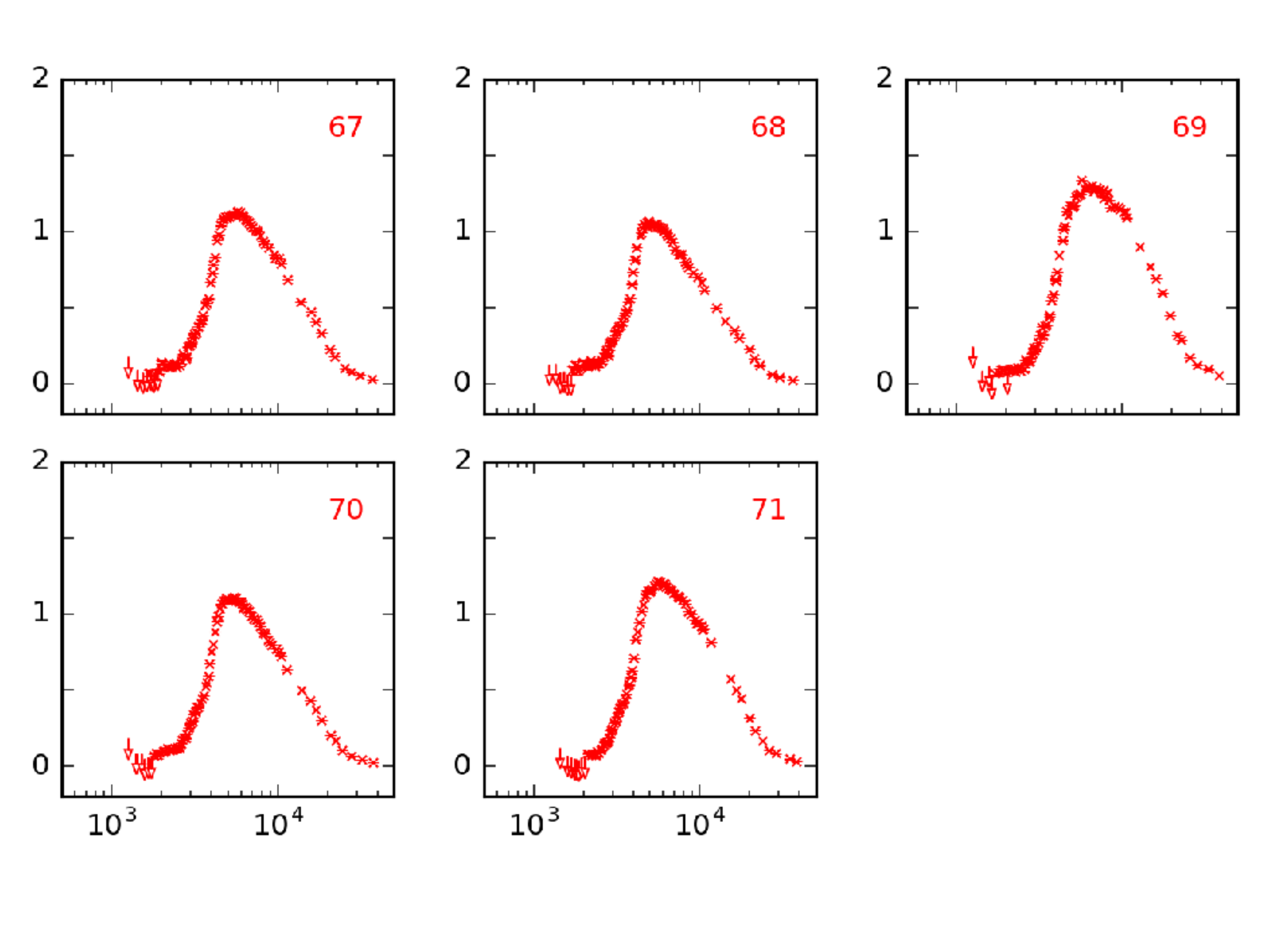}}
	\caption{The set of QG composite SEDs from $1<z<3$ as scaled $F_\lambda$ against wavelength.}
	\label{fig:QG_lo}
	\end{figure}


\end{document}